\documentclass{article}
\usepackage{hyperref}
\usepackage{graphics,graphicx,epsfig}
\usepackage{amsfonts,amsmath,amscd,amssymb}

\begin{document}

\title{Non-perturbative $\lambda\Phi^4$ in $D=1+1$: an example of the constructive quantum field theory approach in a schematic way}

\author{Jorge Gueron$^{\dag,*}$, Mauricio Leston$^{\ddag,**}$}

\maketitle

\vspace{.5cm}

\begin{minipage}{.9\textwidth}\small \it \begin{center}
    $^\dag$ Departamento de Matematica, Facultad de Ciencias F\'isico Matem\'aticas e Ingenier\'ia, Universidad Catolica Argentina \\ Buenos Aires \end{center}
\end{minipage}

\vspace{.5cm}

\begin{minipage}{.9\textwidth}\small \it \begin{center}
    $^\ddag$ Instituto de Astronom\'ia y F\'isica del Espacio, Pabell\'on IAFE-CONICET, Ciudad Universitaria, C.C. 67 Suc. 28, Buenos Aires
     \end{center}
\end{minipage}

\begin{abstract}
During the '70, several relativistic quantum field theory models in $D=1+1$ and also in $D=2+1$ have been constructed in a non-perturbative way. That was done in the so-called {\it constructive quantum field theory} approach, whose main results have been
obtained by a clever use of Euclidean functional methods. Although in the construction of a single
model there are several technical steps, some of them involving long proofs, the constructive quantum field theory approach contains conceptual insights about relativistic quantum field theory that deserved to be known and which are accessible without
entering in technical details. The purpose of this note is to illustrate such insights by providing an oversimplified schematic exposition of the simple case of $\lambda\Phi^4$ (with $m>0$) in $D=1+1$. Because of the absence of ultraviolet divergences in its perturbative version,
this simple example -although does not capture all the difficulties in the constructive quantum field theory approach- allows to stress those difficulties inherent to the non-perturbative definition. We have made an effort in order to avoid several of the
long technical intermediate steps without missing the main ideas and making contact with the usual language of the
perturbative approach.
\end{abstract}

\mbox{}
\raisebox{-4\baselineskip}{%
  \parbox{\textwidth}{\mbox{}\hrulefill\\[-4pt]}}
{\scriptsize$^{*}$ E-mail: jorge$_{-}$gueron@uca.edu.ar}\\
{\scriptsize$^{**}$ E-mail: mauricio@iafe.uba.ar}

\section*{Introduction}

Sometimes we talk about features of theories whose existence has not been proven yet. Such is the case of some statements about non-perturbative phenomena or
the strong coupling regime of a relativistic QFT which is known at the present only in a perturbative way. These statements refer to what is expected if we could go beyond perturbation theory. One non-proved assumption behind such statements is the {\it existence of
the non-perturbative model} from which the perturbative series is derived. Then, it would be desirable to have at hand a concrete simple example of such features in a relativistic quantum field theory defined in non-perturbative way. Such example could be
useful as a toy model illustrating expected non-perturbative phenomena of realistic theories which are known at the present only in a perturbative way.

Actually, there are several examples of interacting relativistic QFT defined in
a non-perturbative way. These models, which were obtained during the '70, include a family of scalar field polynomial interactions, interactions of Yukawa type and also an example of a non-polynomial interaction, both in $D=1+1$ and $D=2+1$. The approach
used for the obtention of these models is known as {\it constructive quantum field theory} (CQFT). See \cite{Summers} for a recent historical account.

In order to see the role of the CQFT approach and its relation with the perturbative approach, let us consider the following diagram:

\begin{equation}
\begin{CD}
\begin{array}{l}
\text{(1) Non-perturbative} \\
\text{meaningless formal expressions}\\
\end{array} @>\text{Taylor expansion}>> \begin{array}{c}\text{(2) Meaningless formal}\\\text{perturbative series}\end{array}\\
@VV\text{Regularization}V     @VV\text{Regularization}V \\
\begin{array}{l}
\text{(1') Non-perturbative} \\
\text{meaningful expressions}
\end{array} @<\text{sum of the series (?)}<<  \text{(2')Meaningful formal series}\\
@>\text{Taylor expansion (?)}>>
\end{CD}\nonumber
\end{equation}

1) In the upper left corner of this diagram, we have the formal expressions for the non-perturbative n-point functions, like those involving functional integrals or the Dyson evolution operator. These expressions are meaningless
for several well known reasons that will be recalled later in this note. However, in the standard exposition these expressions are formally manipulated in order to derive the Taylor series (in power of the coupling constant) in the right upper corner of the diagram.

Then, we should not say that the perturbative series have been {\it deduced} from a more fundamental definition. That misleading conclusion could arise if we take seriously the standard procedure for derivation of the series, ignoring
that the original expression from which the series come (path integral or the Dyson evolution operator) has only a formal meaning. The role of the formal expressions and their manipulations consist in motivating the definition of the perturbative series.

2) In this second step of the diagram, the terms of the series are still meaningless because of the standard ultraviolet divergences. So, the definition of the QFT should begin in a further step. Naively, we can think that it begin after
the regularization procedure in the lower right hand side of the diagram, because there we have a well defined series for the n-point functions of the QFT.

2') However, the QFT is not still defined in this third step, because in order to define the set of n-point functions (which in the end are just numbers) what is needed is a {\it procedure for extracting a number from the series}. But it turns out that most of these perturbative series are {\it divergent} according with the standard convergence criterium. This fact has been addressed from early years of perturbative QFT (see \cite{Jaffediv} for examples of $\lambda\Phi^4$ in $D=2$). Without having a criterium for getting a number from the series, we have not yet defined the QFT \footnote{For particular n-point functions in special theories, we will find in the bibliography efforts to find such criterium, like the Borel convergence. However, such criterium does not hold for the complete set of n-point functions, which should be defined in order to define a QFT model.}.

We want to remark that the issue of the non-convergence of the series is more relevant in the QFT case than in other cases (like non-relativistic quantum mechanics) in which we have a quantity defined by a complicated expression and an asymptotic series is derived in order to make practical computations. In the QFT case, the series are not the Taylor expansion of a non-perturbative well defined n-point function. The perturbative series is all what we have. If we want to avoid the use of euphemisms, we should say that
{\it the perturbative approach does not define a QFT even at weak regime, because the radius of convergence of these series is not small but zero} \footnote{It is true that most of these series are {\it asymptotically convergent} as the coupling constant $\lambda$  goes to zero. The asymptotic convergence ensures that there exist a function of $\lambda$ such that the difference between this function and the truncated series at order $N$ is of order $\lambda^{N+1}$. However, there is not a unique function having this series as its asymptotic expansion. So, the asymptotic convergence is not enough for defining the n-point functions. We will came back to this point later in this note}

1') Now, we can appreciate the achievement of CQFT approach: that corresponds to the process of regularization in the left side of the diagram. We can consider it as a `non-perturbative regularization´. As we will see in this note, most of the models of CQFT have been obtained by making sense of the formal functional integral expression for the n-point functions. The previous considerations show that the achievement of CQFT is not merely a description in strong regime of {\it existent} models described in a weak regime.

Once we defined a non-perturbative expression for the n-point functions, we can make a Taylor expansion and obtain a perturbative series. It could be divergent, but now we should not be so worry, because that series is not supposed to define the theory.

I turns out that in some cases, like $\lambda\Phi^4$ in $D=1+1$, the Taylor expansion of any n-point function coincides with the ordinary perturbative series. In those cases, we can draw the lower arrow from the left to the right, considering that non-perturbative expression arises from a suitable criterium for the resummation of the perturbative series.

Although the models obtained by this procedure are far of being realistic (because they are defined in $D=1+1$ and $D=2+1$), these constitute the first examples of the marriage of special relativity and quantum theory in the interacting case (see the introductory comments in \cite{Jaffe2}). Besides the importance of knowing their existence, we think that certain steps in the construction of these models deserve to be known. That is because in the construction of these models we find certain difficulties which are not present in perturbation theory. These difficulties themselves manifest deep aspects of a relativistic QFT and have a conceptual value. Moreover, by looking at non-perturbative phenomena in a concrete model we could gain intuition about possible non-perturbative effects in QFT models which at the present are known only in a perturbative way.

However, in order to understand the construction of a single model like $\lambda\Phi^4$ in $D=1+1$ (whose construction was initiated in \cite{Jaffe68}) we should face the technical difficulties of several proofs and definitions, which could discourage someone who only want to have a feeling about CQFT. The purpose of this note is to help in this sense, by providing a friendly introduction to the single example of $\lambda\Phi^4$ in $D=1+1$, reducing the amount of technicalities. Concerning the amount of details, this exposition is sited between introductions like \cite{Jaffe2} and more detailed rigourous exposition, full of definitions and theorems, like the self contained book \cite{Jaffe1}. We have written this note in such a way that even an graduate student could have a picture about CQFT. Although there exist good and friendly introductions like \cite{Jaffe2}, which give a general idea without entering in the many definitions and proofs, we think that this note presents new features from the pedagogical
point of view. In the next section, we will describe these features. That section could be skipped, going directly to section {\ref{DefQFT}}.

\section*{The novelty of this exposition}

\subsection*{The audience to whom this notes are written}

The community to whom this note is addressed -the one to which the authors belong- is more comfortable with the use of heuristic arguments, formal manipulation, plausibility considerations. Such is the
kind of language in which we want to communicate the ideas of CQFT. In doing that, we also try to overcome the common prejudice consisting in associating the mathematical style of an exposition of a subject with
the lack of a relevant physical contribution, considering that such exposition contains only a formalization of previous known physical concepts, which could be appreciate in a more fresh and heuristic way.

As we will see, that is not the case of the CQFT approach. In fact, the value of the ideas in the CQFT approach could be appreciated even in a schematic and friendly exposition like this, which contains the following features:

\begin{itemize}
\item plausibility and heuristic arguments instead of rigourous arguments.
\item examples instead of definitions
\item sketches of the proofs instead of proofs
\item inclusion of pictorial descriptions and drawing of parallel with simple examples of the standard approach
\item Restriction to simples cases instead of an exhausted exposition.
\end{itemize}

We are aware that the previous features are scattered in several expositions like the book \cite{Jaffe1}, the early book \cite{Simon}, the textbook \cite{Dimock} or even inside some research papers. However, we can access to these only after avoiding several technical details. This note tries to facilitate the task by collecting this scattered pedagogical tools in a single and short exposition.

We are aware that it is a difficult task to give an accurate exposition of CQFT using exclusively heuristic considerations, avoiding several definitions and proofs. In fact, the achievement of CQFT was precisely that of showing that interacting QFT models exist as
mathematically well defined objects. It is natural to ask what is left -besides an historical exposition- if we omit that aspect of the construction. It is not easy the decision about
which part of the rigorous procedure could be cut without turning the exposition a mere vulgarization of the subject.

However, we hope that this oversimplified exposition -which keeps selected intermediate inequalities- could still give a feeling about which are the specific difficulties towards the non-perturbative definition which we never confront in the perturbative approach

One last comment on the purposes of this note: it does not intend to be a historical account of CQFT. For that purpose, a good and recent reference is \cite{Summers}. In fact, our pedagogical aims conduct to restrict the consideration on the single example of $\Lambda\Phi^4$. Accordingly, we have quoted a very short list of references.
Even though the example of this note does not capture all the difficulties involved in CQFT, it is enough rich for a first approach.

\subsection*{Organization of the material}

We have organized the article in several parts.

In part I, we briefly describe what a relativistic quantum field theory, the constructive approach is attempting to construct, is. In particular, in section \ref{inbrief}, we make a brief summary of the difficulties for the definition of $\lambda\Phi^4$ model in $D=1+1$. Those difficulties will be considered in more detail along the note.

Part II introduces the statistical description of quantum mechanics in terms of Gaussian processes, starting from the simple case of the anharmonic oscillator and going to free QFT in $D=1+1$

Part III is the main part of the note. There, in sections \ref{dospuntosSection}, \ref{StabilitySection} and \ref{volumeninfinito}, it is described schematically how we can deal with three different kind of possible divergences in order to properly define the Euclidean n-point functions. It concludes with the description of the steps for the proof that these well defined n-point functions fulfill the desired physical requirements.

Part IV is devoted to link the exposition of the previous part (which is done in the Euclidean functional approach) with the non-perturbative Hamiltonian approach and also with the usual perturbative approach. This last part is important in
order to see that the models of CQFT are not merely an abstract construction but the materialization of ordinary notions.

We end with some remarks and a guide to a further reading.

\tableofcontents
\newpage

\part{Definition of a relativistic QFT and the constructive strategy}

\newpage
\section{General properties defining a relativistic QFT}\label{DefQFT}

Before considering any strategy, we should know what a relativistic QFT we want to construct is. If the standard
quantization procedure -which is the one used for the free QFT- were not plagued of difficulties and ambiguities when we add interactions, the question about the definition of a relativistic QFT could
be considered an unnecessary worry. We could simple state: ``a relativistic QFT is a quantum theory obtained by the canonical quantization procedure applied to a relativistic classical field theory". However, as we will see,
such canonical procedure is difficult to implement for an interacting classical field theory.

\subsection*{Difficulties for the definition of the interacting term}
The difficulties arise from the very beginning: {\it the definition of the quantum interacting term}. In quantum mechanics, there is not such a problem in this step. In the Heisenberg representation, $\hat{x}_{(.)}$ is an operator valued {\it function} of the time: it takes
a real number $t$ and gives an operator as an output. So, there is not a big problem in defining an interaction term as a suitable function of the position operator. That fact makes possible the existence of quantum mechanics systems with
a non-free Hamiltonian\footnote{We are aware of the following fact: because the
position operator
is an {\it unbounded operator}, the domain of the power of the position operator could not coincide with the one of $\hat{x}_{(.)}$. So, is not straightforward a proper definition of the interacting term. However, an appropriated restriction of the domain allows to define
rigorously the interacting term as a function of the position operator. In the QFT case, we will find more difficulties}

In the QFT case, $\hat{\Phi}(.)$ is an operator valued {\it functional}, which takes a {\it function of the spacetime} and gives an operator as an output. The notation $\hat{\Phi}(x)$, which suggests that $\hat{\Phi}$ is an operator valued function,
comes from an abuse of language. It comes from the existence of special types of linear distributions $H$ (let us call them {\it regular} distributions) such that their action on a function $f$ (the test function) can be written as an integral: $H(f)=\int h(x) f(x) dx$, being
$h$ a function called {\it the kernel} of $H$. A non dangerous abuse of language, which simplifies the notation, consists in denoting the kernel with the same letter than the one used for the distribution. It allows to write: $H(f)=\int H(x)f(x) dx$. $H$ stands
for a distribution in the l.h.s and for a function in the r.h.s.

However, a second abuse of language, a bit more dangerous than the previous one, consists in extending the previous integral expression to distributions which does not admit a kernel. Such is the case of $\hat{\Phi}(.)$ which is not a regular distribution. When the expression $\hat{\Phi}(x)$ is
written, $\hat{\Phi}$ is implicitly considered as a regular distribution admitting an integral representation, in which $\hat{\Phi}(x)$ is smeared with the test function $f$:

\begin{equation}
\hat{\Phi}(f)=\int \hat{\Phi}(x)f(x)dx\label{abuse}
\end{equation}

A similar abuse of language with the distribution $\delta_{x_0}$ defined by $\delta_{x_0}(f)\equiv{f(x_0)}$. The abuse of language consists in writing the action of this functional as: $\delta_{x_0}(f)=\int f(x)\delta(x-x_0)dx$.
Here,  $\delta(x-x_0)$ is considered formally as a function, which is supposed to be the kernel of $\delta_{x_0}$. However, such a kernel does not exist.

The absence of a kernel for the distribution $\hat{\Phi}$ -hidden in the previous abuse of language- is the root of the difficulties for the definition of the interacting term in the equation of motion of the field. Let us remark that e.o.m for the free
scalar field, when is written properly in a distributional way, is:

\begin{equation}
\hat{\Phi}(\square f + m^2 f)=0
\end{equation}

By writing the formal expression $\hat{\Phi}(f)=\int \hat{\Phi}(x)f(x)dx$ and doing an integration by parts, we arrive at the usual form of the e.o.m: $(\square + m^2)\hat{\Phi}(x)=0$. Accordingly, the wished interacting e.o.m should be something like:

\begin{equation}
\hat{\Phi}(\square f + m^2 f) + \hat{R}(f)=0
\end{equation}

where $\hat{R}$ is a linear distribution, constructed in terms of the original $\hat{\Phi}$.

However, it turns out that for the quantum field distribution such a term is difficult to construct. When a distribution $H$ admits a kernel $h$, we can perform operations on $H$ by an analogous manipulation of its kernel. For instance,
we can define $H^4$ as a distribution whose kernel is $h^4$ (of course, after taking into account suitable restrictions on the space of the test functions). Because $\hat{\Phi}$ is not a regular distribution admitting a kernel, it does not make
sense to define ${\hat{\Phi}}^4(.)$ as the distribution given by: ${\hat{\Phi}}^4(f)=\int{\hat{\Phi}}^4(x)f(x)dx$. That is the beginning of all the difficulties which will be considered in this note.

It is important to remark that the previous difficult is not related to the fact that $\hat{\Phi}(f)$ is an unbounded operator. The unboundedness of $\hat{\Phi}(f)$ leads us to make appropriated restriction on
the domain in order to make sense of composition of operators like: ${(\hat{\Phi}(f))}^4$. However, the last expression is different from the one needed for the definition of the interaction term $R$ because ${(\hat{\Phi}(.))}^4$ {\it is not a a linear functional}. When in
standard textbooks is written the interacting e.o.m containing the additional term  $:\hat\Phi^3(x):$, it is implicit that we should read this equation in a functional sense, considering that $:\hat\Phi^3(x):$ is the formal kernel of a linear distribution. We made this remark in order to stress that the difficulties behind the definition of the interaction term are not of the same nature than the ones which we find in ordinary quantum mechanics.

\subsection*{Garding-Wightman axioms: minimal features of a free field that should be valid in any relativistic QFT}
The previous observations lead us to define an interacting QFT not as the output of a
procedure difficult to be implemented in general but by requiring those minimal features of a free theory that we want to keep even in presence of interaction. These features are:

 \begin{itemize}

 \item Poincare covariance of the field operators
 \item Existence and uniqueness of the vacuum (an state invariant under the Poincare action)
 \item The inclusion of spectrum of the momentum operator in the forward light cone.
 \item Commutation of the fields $[\hat{\Phi}(x),\hat{\Phi}(x')]=0$ for points $x,x'$ spacelike separated\footnote{Anti-conmutation relation for fields of half-integer spin. The equality $[\hat{\Phi}(x),\hat{\Phi}(x')]=0$ is a short version of the proper statement: $[\hat{\Phi}(f),\hat{\Phi}(g)]=0$ is the support of $f$ and $g$ are spacelike separated. We can read this from the previous formal statement by smearing both sides of the formal equality with $f(x)g(x')$ and integrating over $x$ and $x'$. This is the type of abuse of language that we have mentioned before, which does not conduct to any wrong statement if it used within certain limit. We will make use of this practical language several times along this note}
 \end{itemize}

This minimal selection of properties of the free field was expressed in a precise and organized way by Arthur Wightman in the middle of '50, soon after all the ingredients of perturbation theory were introduced. These properties are known as {\it Garding-Wightman (GW) axioms} \cite{Wigthman}.

Due to common prejudices, the term {\it axioms}, when used in an exposition of a physic theory, is associated to requirements motivated by aesthetical or mathematical reason which go
beyond physical considerations. We want to emphasize that GW axioms intend to capture physical features of the free field which any QFT theorist would hardly abandon. For that reason, we prefer to talk about {\it general properties of a relativistic QFT} instead of GW axioms.

Going back to the original question, we can say that a relativistic QFT consists of a Hilbert space and a family of operators fulfilling certain physical requirements
encoded in the GW axioms. Although it is easy  to understand the statement of each of GW axioms, the task of providing examples fulfilling these axioms is very difficult. The first non trivial example was obtained during the end of '60, as result of a remarkable work of Arthur Jaffe and James Glimm in a sequence of papers starting in \cite{Jaffe68}. That was done by using an
approach close to the canonical quantization, consisting in defining the interacting Hamiltonian in order to define the interacting field operator. Although this approach has the advantage of admitting a more clear physical interpretation, we will consider another
equivalent approach that was developed later, which was more suitable for practical purposes.

\subsection*{Reconstruction of QFT from the set of Poincare invariant n-point functions}\label{DefQFT}

An important step towards a useful definition of a general relativistic QFT was the investigation of the properties of the vacuum correlation functions $W_n(x_1,x_2,...,x_n)\equiv(\Omega^{int},\Phi(x_1)\Phi(x_1)...\Phi(x_n)\Omega^{int})$, being $\Omega^{int}$ the
vacuum of the interacting theory. These are the so-called {\it n-point functions}. From the GW axioms it is possible to derive properties of these vacuum correlation functions in a general QFT. It is natural
to ask about the inverse problem: if we have an infinite set of n-point functions, {\it how can we know if these are the vacuum correlation of a QFT fulfilling the GW axioms?}

The answer is given by {\it Wightman reconstruction theorem} \cite{Wigthman}. That theorem is a consequence of the identification of those properties of a set of all n-point functions $W_n(x_1,x_2,...x_n)$ (for any $n$) which are enough to guarantee that these are the vacuum correlation functions of a relativistic QFT. Omitting the technical details and restricting to the case of scalar field, these properties are:

\begin{itemize}
 \item Poincare invariance.
 \item Symmetry under permutation of two spacelike separated arguments.
 \item Spectral properties (coming from the inclusion of spectrum of the momentum operator in the forward light cone)
 \item Positivity condition: a set of inequalities expressing that the n-point functions come from an inner product in the Hilbert space.
 \item Cluster property: factorization of the n-point functions when the spacelike separation of two arguments becomes large. This is related to the uniqueness of the vacuum
\end{itemize}

These requirements, together with technical requirements on the smoothness of these functions, are known as {\it Wightman axioms} for the n-point functions. The proof of the reconstruction theorem is by providing the procedure for the reconstruction of the Hilbert space and the field operators fulfilling the GW axioms.

We have referred to $W_n$ as functions. This is again an abuse of language, that considers $W_n$ as a kernel of the distribution $\tilde{W}_n$. Formally, we can write $\tilde{W}_n(f)=\int{W}_n(x_1,....x_{Dn})f(x_1,....x_{Dn}) (d^Dx)^n$. We should not read
this expression literally, because the distribution $\tilde{W}_n$ does not admit a function $W_n$ such that the previous expression makes sense. This abuse of language is similar to the one made in Eq. [\ref{abuse}]. Just for practical purposes, we will still talk about the n-point ``functions" (and we will also drop the tilde for the distribution).

\subsection*{Reconstruction of QFT from a set of Euclidean invariant n-point functions}

Yet in the early years of pertubative QFT, J.Schwinger had emphasized the practical value of using an imaginary time version of the vacuum correlation functions. It was shown by A.Wightman that time ordered n-point functions fulfilling GW axioms can be analytically continued to imaginary
 time. Formally, that extension corresponds to the change $t\rightarrow{it}$: the so-called {\it Wick rotation}. The development during the '60 of the Euclidean version of the vacuum correlation functions lead to a formulation of an Euclidean version of the Wightman reconstruction theorem. That was done by Osterwalder and Schrader in 1973 \cite{OS}.

What they found was another set of requirements -called the Osterwalder-Schrader (OS) axioms- to be obeyed by the analytic continuation of the n-point functions. In order to recover the Wightman functions, essentially what we have to do is a Wick rotation ($t\rightarrow{it}$). Due to the early role of Schwinger in this issue, the n-point functions fulfilling the (OS) axioms are known as {\it Schwinger functions}.

Omitting technical requirements and restricting to the  case of interest for a scalar QFT, the requirements of the OS axioms for the Schwinger functions are the following:

\begin{itemize}
 \item Euclidean invariance
 \item Symmetry under any permutation of its arguments (the Euclidean translation of the second property of the Wightman functions)
 \item Reflection positivity: a set of inequalities analogous  to the positivity condition, involving reflection of the time in some arguments
 \item Cluster property: factorization of the n-point functions when two arguments becomes large (in the Euclidean sense) separated.
\end{itemize}

In the OS axioms, there is not a requirement analogous to the spectral condition. In this sense, the OS axioms seem to be more economical. There are several subtleties that we are omitting in this schematic presentation. One of these is related to the equivalence of the OS and W axioms. In the original paper where these axioms were formulated \cite{OS} we can find comments about modifications of these axioms which result to be more useful for the construction of models although stronger than the W axioms.

\section{A non-perturbative quantization procedure}\label{Estrategia}

 After having settled the problem -what a generic relativistic QFT is- we can now consider the strategy for its construction. The question is: {\it how would we get a guess for the Schwinger or Wightman n-point functions?}. It could be desirable to have an ansatz for the Schwinger function, having a chance of being successful. Even more, it would be nice to have a quantization procedure. That could be the case if the chosen ansatz were dictated by a classical field theory. That ansatz exists
 and comes from a combination of {\it Feymann-Kac} and {\it Gell-Mann-Low} formulas. The first one establishes a link between Gaussian processes and quantum mechanics.

\subsection{Free QFT in terms of Gaussian process and well defined path integral}

There exists a close connection between expectation values in certain quantum mechanics system (i.e., a non relativistic particle under certain class of potential) and statistical
expectations of {\it Gaussian processes}. That relation is encoded in the so-called {\it Feynman-Kac} formula, which allows to express   $<x'\mid e^{-tH}\mid{x}>$ (with $t>0$, and $H$ being the Hamiltonian) as the expectation of a
Gaussian process. Let us notice the difference in the meaning of the word ``expectation'' in each side of the relation: in the first case, it refers to an inner product of two states. In the second one, it is the mean of products of random variables having a Gaussian probability distribution. Because of that, the previous formula makes the connection between quantum and statistic mechanics possible .

It turns out that the expectation value of the Gaussian process admits a functional integration description, as an integral over a set of path starting at $x$ and finishing at $x'$ weighted with
certain measure. We want to emphasize that what admits a path integral representation is not $<x'\mid e^{itH}\mid{x}>$ but its analytic continuation, which roughly speaking consists on replacing $t$ by $it$.

This relation holds also in the free quantum field theory case: the n-point function, after the Wick rotation, can be written as a path integral of products of Gaussian processes. The correspondent measure
is Euclidean invariant. We see here the existence of a close relation among several different things: relativistic QFT, statistical mechanics, Euclidean invariance and functional integration.

\subsection{The heuristic role of the Gell-Mann-Low formula}

 A more useful formula arise when we combine Feynman-Kac formula with the so-called {\it Gell-Mann-Low} formula, which can be proven in certain quantum mechanics systems. In the case of an anharmonic quantum mechanics system, Gell-Mann-Low formula expresses interacting correlation functions of the interacting position operators in terms of free vacuum correlation functions of position operator of the harmonic case. The terms 'interacting'( 'non-interacting') refers to the anharmonic (harmonic) case. This formula arises after expressing the free vacuum state in terms of the interacting vacuum. By combining Gell-Mann-Low with Feynman-Kac formula, we can express the analytic extension of the interacting vacuum correlation function in terms of expectation values of Gaussian processes. We will still call Gell-Mann-Low to such combined formula.

By {\it formal manipulations}, this version of Gell-Mann-Low formula formula can be extended to the case of an interacting scalar QFT. That is:

\begin{equation}
\boxed{
(\Omega^{int}, \hat{\Phi}^{int}_{it_1}(x_1)...\hat{\Phi}^{int}_{it_n}(x_n)\Omega^{int})=
\lim_{T\rightarrow\infty}\frac{E(\Phi_{t_1}(x_1)...\Phi_{t_n}(x_n)e^{-\int_{-T}^{T}V(\Phi_t)dt})}{E(e^{-\int_{-T}^{T}V(\Phi_t)dt})}
}\label{GL}
\end{equation}

The meaning of the left hand side is clear: it is the (imaginary time version of) the vacuum expectation value of $n$ products of field operators at different instant (chronologically ordered), being $\Omega^{int}$ the vacuum of the interacting theory; the $x_i$'s stand for the spacial coordinates.

The r.h.s has a completely different meaning: it is the mean (expectation), denoted by $E$,  of a product of Gaussian processes -denoted by $\Phi_t$ without the hat- weighted by the exponential factor. The exponent contains the interacting potential $V(\Phi_t)$, which is a spatial integral of a density. This expectation can be formally written as a functional integral with respect to a Gaussian measure $d\mu$. The exponential factor is the one which perturbs the Gaussian measure.

Has we have said, in certain quantum mechanics system - in which we have no spatial coordinates as arguments- it is a rigourous equality and the expectation of the r.h.s can be expressed as a functional integral, which is rigourously defined. In the interesting case of an interacting QFT, {\it the Gell-Mann-Low formula should be considered as an ansatz for Schwinger n-point functions}. In case
that we could prove that the n-point functions obtained in this way fulfill the OS axioms, we can obtain (via the reconstruction theorem) the interacting correlation function fulfilling the Wightman axioms.

As we will see, this ansatz has a high chance to fulfill the OS axioms. By using this ansatz it has been obtained several models fulfilling the Wightman axioms in the '70.

The previous strategy is summarized in the following diagram:

\begin{equation}
\begin{CD}
\begin{array}{l}
\text{{\bf Classical field theory}}\\
\text{{\bf with an interacting}}\\
\text{{\bf term $S_{int}$}}\end{array}
 @>>> \begin{array}{l}
 \text{{\bf Gell-Mann-Low}}\\
  \text{{\bf ansatz}}\end{array}@>\text{Proof of}>\text{OS axioms}> \text{{\bf Schwinger functions}} \\
@V\text{Quantum}V\text{version?}V @. @V\text{Wick rotation}VV \\
\text{{\bf Quantum field eq.}} @<<< \text{{\bf Relativistic QFT}} @<<\text{Reconstruction}< \text{{\bf Wightman functions}}
\end{CD}\nonumber
\end{equation}

So, by using a Gell-Mann-Low formula, one follows a quantization procedure, which takes a classical theory (given by the interacting term which perturbs the free Gaussian measure) and end up with a
quantum theory. We want to emphasize that there is not guaranteed that the resulting quantum theory will exhibit a close relation to the classical theory used as a seed. For instance, it is not guaranteed that the interacting field will fulfill the operator version of the classical equations of motion.

\section{The $\lambda\Phi^4$ in brief}\label{inbrief}

Let us anticipate the steps that we should follow in order to construct the QFT corresponding to $\lambda\Phi^4$. Those are summarized in the following table and can be organized in two parts:
I. Dealing with divergences in order to define the candidates to be the Schwinger functions. II. Verifying that these functions fulfill all the physical requirements (OS axioms).

\subsection{I. Dealing with three types of divergences}

In this first part we should prove that the r.h.s of the Gell-Mann-Low formula is well defined. It is not trivial because there are three different divergences that could arise:
\begin{enumerate}
\item {\it Definition of the interaction term}

As we have mentioned, in spite of the so-called operator fields are operator valued {\it distributions} (not just operator valued {\it functions}) which do not admit kernel, the expression $\hat{\Phi}^4(x)$ does not make sense. However, we can consider a family of functions $h_{\kappa}^{(x)}$, labeled by an integer number $\kappa$,
localized (in a precise sense to be specified later) around $x$, such that as $\kappa\rightarrow\infty$, $h_{\kappa}^{(x)}$ approaches -in a sense of distributions- to the $\delta_x$. Then, we can consider an operator valued distribution
$:\hat\Phi_{\kappa}^4:$, indexed by $\kappa$, whose kernel is defined by $:\hat\Phi_{\kappa}^4:(x)\equiv\;:{(\hat{\Phi}(h_{\kappa}^{(x)}))}^4:$. Here ''$:\;:$'' stands for the Wick order. It is important to remark that $:\hat\Phi_{\kappa}^4:(x)$ is supposed to be the kernel of the distribution $:\hat\Phi_{\kappa}^4:$. It means that the action of it on a function $f$ is defined by
$:\hat\Phi_{\kappa}^4:(f)=\int :\hat\Phi_{\kappa}^4:(x) f(x) d^2x$.

In the Gaussian process description, the Wick order has a counter part which will be also denoted as ``$:\;:$''.
As we will see, the result of applying $:\;:$ to the $n$-power of a Gaussian process is defined as a polynomial of the same degree $n$ but with the addition of lower powers. The usual Wick order of a product of field operator can be written in the same way; it is a more convenient reformulation of the usual operation consisting in moving the annihilation operator to the left. As we will see, this $\kappa$ dependent Gaussian process $:\Phi_{\kappa}^4:$ can be expressed as:

\begin{equation}
:\Phi^4_{\kappa}:(x)\equiv(\Phi_{\kappa}(x))^4-6c_{\kappa}(\Phi_{\kappa}(x))^2+3(c_{\kappa})^2
\end{equation}

where the coefficient $c_{\kappa}$ is the expectation value: $c_{\kappa}\equiv{E}({\Phi_{\kappa}(x)}^2)$.

The distribution needed for the definition of the interacting term is defined as the limit $\kappa\rightarrow\infty$ of the previous expression. In this limit the coefficients $c_{\kappa}$ diverges; also, both  $(\Phi_{\kappa}(x))^4$ and $(\Phi_{\kappa}(x))^2$ diverges in a sense of distribution.
However, the limit of the total sum in the r.h.s exists in $D=1+1$. The limiting distribution is we we call  $:\Phi^4:$\footnote{We want to call the attention to the notation that we are following: we write $:\Phi^4:(x)$ instead of the usual $:\Phi^4(x):$.
This notation emphasizes that $:\Phi^4:$ is a distribution being $:\Phi^4:(x)$ its (formal) kernel. Accordingly, when we evaluate this distribution on a function $f$, we can write the formal expression: $:\Phi^4:(f)=\int:\Phi^4:(x)f(x)d^2x$.}. This limit will be described in more detail in Section \ref{dospuntosSection}.

In order to define the interacting term appearing in the exponential of the r.h.s of the Gell-Mann-Low formula, we should evaluate the distribution $:\Phi^4:$ in a suitable function $f$ of $R^2$. Let us consider a function $f$ with compact support in a bounded region of the spacetime  $\Lambda$ having the value $\lambda$ in this region.
So, we can write the interacting term $\frak{A}^{\Lambda}\equiv\;:\Phi^4:(f)$ as:
$\lambda\int_{\Lambda}:\Phi^4:(x)d^2x$. This is the cut-off interacting term.

\item {\it The quantum boundedness of the interacting term}

Having defined the cut-off interacting term, it should be proved that the r.h.s side of the Gell-Mann-Low formula is well defined. It requires:

\begin{equation}
\int e^{-\frak{A}^{\Lambda}} d\mu_{gaussian} < \infty
\end{equation}

Of course, it is also needed the convergence of the $\int (..) e^{-\frak{A}^{\Lambda}} d\mu_{gaussian}$, where the dots refer to any products of Gaussian fields. Although we have started with a polynomial bounded from below,
the convergence of the above integral is not guaranteed because the Wick product has destroyed that boundedness. However, it will be shown in section \ref{StabilitySection} that this integral converges.
\item {\it The removal of the cutoff: the infinite volume limit}

The removal of the cut-off requires that the limit $\Lambda\nearrow{R}^2$ exists. For general interactions, more complicated than this example, the so-called {\it cluster expansion} is used in order to prove that the infinite volume limits for the Schwinger functions exists. In particular,

\begin{equation}
\lim_{\Lambda\nearrow{R}^2}\text{Schwinger Functions}_{\Lambda}< \infty
\end{equation}

More details in Section \ref{volumeninfinito}

\end{enumerate}

\subsection{II. Verifying that this n-point functions comes from a RQFT}

The steps of part I are required in order to show that the n-point functions exist. However, after accomplishing these steps, it remains to be proved that these n-point functions fulfill the general properties encoded in the OS axioms. The nice feature of the Gell-Mann-Low ansatz is
that most of the properties axioms are fulfilled in the cut-off n-point functions, and these properties are preserved under the infinite volume limit. We can decompose the OS axioms in two groups:

\begin{enumerate}
\item Euclidean invariance, reflection positivity,symmetry
\item Cluster property and regularity
\end{enumerate}

In the first group we have the Euclidean invariance (which formally can be checked when the infinite volume is taken), the symmetry, which is the counterpart of the locality in the Minkowski case and is manifest in the ansatz and reflection positive, which expresses that the n-point functions come from an inner product. The last one is not difficult to be checked in the cut-off version.

The more difficult part is the proof of the cluster property, which is not manifested in the ansatz. The regularity conditions refer to smoothness properties of the n-point function and they will not be considered in this introductory note.

\vspace{10 mm}

{\center
\begin{tabular}{|l|c|}
  \hline
  {\bf I Dealing with divergences} & \\
  \hline
  \hline
  {\it Definition of the interacting density} & Existence of the limit\\
$\lambda:\Phi^4:(x)$ &
$\lim_{\kappa\rightarrow\infty}{\Phi_{\kappa}(x)}^4-6c_{\kappa}(\Phi_{\kappa}(x))^2+3{c_{\kappa}}^2$ \\
& \\
  \hline
  {\it The quantum boundedness} & \\
   {\it of the cut-off interacting term}&
$\int e^{-\frak{A}^{\Lambda}} d\mu_{Gaussian} < \infty$
 \\
$\frak{A}^{\Lambda}\equiv\int_{\Lambda}:\Phi^4:(x)d^2x$ & \\
& \\
  \hline
  {\it The removal of the $\Lambda$ cutoff} & \\
  & $\lim_{\Lambda\nearrow{R}^2}\text{Schwinger Functions}_{\Lambda}< \infty$\\
   &  \\
  \hline
  \hline
  {\bf II Verifying Axioms} & \\
  \hline
  \hline

  {\it Euclidean Invariance} & \\
  {\it Symmetry} & Almost trivial. \\
  {\it Reflection Positivity} & Manifested in the ansatz \\
  & \\
  \hline
  {\it Clustering} & Hardest part of the proof. \\
  & \\
  \hline

  {\it Mass Gap}: & This proof is related\\
  {\it allows particle interpretation}& to the one of clustering \\
  & \\
  \hline
\end{tabular}
}

.

As we have anticipated, we will not provide the technical detail of all the steps but a oversimplified version. The omission of the intermediate steps will be more important in the most difficult part: the infinite volume limit
and the proof of the cluster property.

We can see in the table an additional requirement, the {\it mass gap}, which is not included in the OS axioms. This is a sufficient criterium for the existence of a particle interpretation.

\newpage

\part{The link between quantum mechanics and probability: the Wick rotation}

\newpage

\section{Interacting QFT in D=1 revisited}\label{InteractingQM}
In this section we will reconsider the quantum mechanics of harmonic and anharmonic oscillators (a QFT in $D=1$) using a Gaussian process description which will be useful for the case $D=1+1$. The reader could take a look at \cite{Dimock} for a friendly and more detailed exposition.

\subsection{Gaussian processes and path integral}

\subsubsection{Generalities on Gaussian variables}

Instead of providing a general definition of a {\it random variable}, we will start with the case of a {\it Gaussian variable}, which is the only relevant for our purpose. In order to define it, what we need is just a space $M$, certain subset of it,
representing the possible outcomes, and a measure $\mu$ which assigns  probability to the different outcomes. A random variable is real valued function $\Phi$ on $M$. We can compute the probability that
$\Phi$ takes its values in some interval $B$ as follows:

\begin{equation}
Prob (\Phi\in{B})={\mu}(\Phi^{-1}(B))
\end{equation}

This definition makes sense if the pre-imagine of $B$ by $\Phi^{-1}$ is one of the subset of $M$ to which we can assign probability. We define the {\it expectation or mean}, denoted by $E$, of any function\footnote{Indeed, not any function. The function should be such that $F(\Phi)$ wil be again a random variable. We will not enter in the statement of this condition} $F(\Phi)\equiv{F}\circ\Phi$ (being $F$  a real valued function $F:R\rightarrow{R}$) as

\begin{equation}
E(F(\Phi))\equiv\int_{M}{F}(\Phi){d}\mu
\end{equation}

In particular, a {\it Gaussian variable} $\Phi$ of mean $a$ and {\it covariance} $C$ is a random variable such that for any function $F$ the expectation value is:

\begin{equation}
E(F(\Phi))=\frac{1}{(2\pi{C})^{\frac{1}{2}}}\int F(x)e^{-\frac{(x-a)^2}{2C}}dx
\end{equation}

We see here how we can express the expectation value as an ordinary integral of the real function $F$ weighted by an exponential.

From the definition, it is clear that the numbers $a$ and $C$ are the following expectation values:

\begin{equation}
a=E(\Phi)\;\;\; C=E((\Phi-a)(\Phi-a))
\end{equation}

The important property of the Gaussian variables is that these two expectation values determine all the remaining expectation values. We will restrict to the case $a=0$.

In order to make contact with quantum mechanics and QFT we need more than a single Gaussian variable. Let us first consider the case of a finite number of Gaussian variables. It is said that the set of variables $\Phi_{i}$ (with $i=1,...n$) are {\it jointly Gaussian} if there exist a positive-definite matrix $C$, the {\it covariance matrix},
such that for any set of functions $F_i$, the expectation value of $\prod_{i=1}^{i=n}F_i(\Phi_i)$ is:

\begin{equation}
E(\prod_{i=1}^{n}F_i(\Phi_i))=\int\prod_{i=1}^{n}F_i(x_i)d\mu_n\label{integralmulti}
\end{equation}

being  $d\mu_n$ the following measure:
\begin{equation}
d\mu_n=  \frac{1}{(2\pi)^{\frac{n}{2}}(Det C)^{\frac{1}{2}}} e^{-\frac{1}{2}x^TC^{-1}x}\prod_{i=1}^{n}dx_i\label{medidagaussiana}
\end{equation}

As in the case of a single Gaussian variable, the coefficients of the matrix $C_{ij}$ are the expectations:

\begin{equation}
C_{ij}=E(\Phi_i\Phi_j)
\end{equation}

Now, let us consider an infinite set $I$ of indexes and an infinite dimensional matrix $C$ with the property that each finite dimensional $n\times{n}$ block arising from restricting the set indexes to finite subsets of $I$ $\{t_1,t_2,...t_n\}$ is positive-definite.

A theorem due to Kolmogorov (see \cite{Dimock}, Theo 11.11) assures that under the previous conditions there exist a collection of (infinities) random variables indexed by $I$ called {\it Gaussian process} such that for each finite choice of the indexes $\{t_1,...t_n\}$ the corresponding random variables are jointly Gaussian variables having as covariance the $n\times{n}$ matrix $C$.\footnote{The theorem says something more general. However, we are interested here in this particular consequence}.

For the quantum mechanics case, it will be relevant the case in which the set of indexes $I$ is a real interval. We can consider that $\Phi_t$ describes a random walk of a particle, being one of the random value of $\Phi_t$ the position of the particle at the instant $t$.

\subsection{Path integral representation}

Until now we have shown how to compute the expectation of functions of a special type: those whose (random) values are determined by a finite set of values $\{\Phi_{t_1}...\Phi_{t_n}\}$. Considering $\Phi_t$ as describing a random walk, it could seem that we are able to compute only the probability of the event defined by a finite number of outcomes; i.e., the probability of finding the particle in certain range of values at a finite set of instants $\{t_1,t_2,...,t_n\}$. However, the existence of the Gaussian process that follows from  the Kolmogorov theorem means that it should make sense to assign probability to other outcomes, like the outcome:{\it the trajectory is contained in certain range of paths}. In the particular case of a Gaussian process, that in turn allows us to compute the expectation of expressions of the type $F[\{\Phi_t\}]\equiv\int_{a}^b \Phi_t dt$. This expression depend on the whole history of the Gaussian
process along the interval $[a,b]$, and not only on its values at a finite number of instants.

Then, for such generalized functions, we can still write:

\begin{equation}
E(F[\{\Phi_t\}])=\int F[\{\Phi_t\}]d\mu
\end{equation}

It is important to emphasize that the previous expression, although well defined, does not admit a simple integral representation like the one in Eq. \ref{integralmulti} because it does not make sense the limit $n\rightarrow\infty$ for $d\mu_n$ in  Eq. [\ref{medidagaussiana}]. Such would lead to the following  meaningless expression (which has only an heuristic value):\\

\begin{equation}
E(F[\{\Phi_t\}])=N \int F[x(t)] e^{x[t]C^{-1}x[t]}Dx[t]
 \end{equation}

In this formal statement, $N$ is an infinite normalization constant and the integral is over all the paths $x[t]$. Because such expression has only a formal meaning, it can not be used for the derivation of furthers theorems and properties. Sometimes in the
literature we see how such meaningless expressions are manipulated in order to get results in a direct way. An example of that is the obtention of the Feynman rules in the functional approach. Although the results are legitimated by other rigorous means, the
formal manipulation of the path integral has an heuristic value which justifies its use.

\subsection{Oscillator process and quantum harmonic oscillator}

In certain quantum systems, like the harmonic oscillator and relativistic free fields, we can find a natural positive- definite matrix which guaranties the reconstruction of a Gaussian process. As we will see, this matrix will be related to a vacuum correlation function.

\subsubsection{Schwinger two-point functions of the harmonic oscillator}
Let us consider a quantum 1-dimensional harmonic oscillator of unit mass and frequency $m>0$  and let us denote by $\Phi_t$ the position operator at time $t$ in the Heisenberg
representation. The last one can be expressed as: $\hat{\Phi}_t=e^{iH_0t}\hat{\Phi}_0e^{-iH_0t}$, being $\hat{\Phi}_0$ the position operator at $t=0$. We have redefined $H_0$ in such a way that $H_0\Omega=0$, being $\Omega$ the vacuum.  This operator fulfills an equation of Klein-Gordon type in $D=1$:

\begin{equation}
(\partial^2_t+m^2)\hat{\Phi}_t=0
\end{equation}

We can consider this as the $D=1$ version of the Klein-Gordon field operator.

Let us consider the vacuum correlation function of the products of the field at two different instants $W_2(t_1,t_2)\equiv(\Omega,\hat{\Phi}_{t_1}\hat{\Phi}_{t_2}\Omega)$, being $\Omega$ the vacuum. This vacuum correlation (which is function of the 2-instants) gives the transition amplitude between an eigenvector of the Hamiltonian
and itself after time evolution. Then, this quantity will be
just a phase. This correlation function can be written as:

\begin{equation}
W_2(t_2,t_1)=(\Omega,\hat{\Phi}_0e^{iH(t_2-t_1)}\hat{\Phi}_0\Omega)=\frac{e^{im(t_2-t_1)}}{2m}
\end{equation}

What has a chance of being a covariance of a Gaussian process is not this quantity but its imaginary time extension $S(t_1,t_2)\equiv{W}_2(it_1,it_2)$ for $t_1\leq{t_2}$. Using the following useful relation

\begin{equation}
\int \frac{e^{ips}}{p^2+m^2}dp=\pi\frac{e^{-m\mid{s}\mid}}{m}\; (m>0).
\end{equation}

we can attempt to define a covariance matrix $S_2(t_1;t_2)$ for $t_1\leq{t_2}$ by:

\begin{equation}
S_2(t_1;t_2)=\frac{1}{2\pi}\int\frac{e^{ip(t_2-t_1)}}{p^2+m^2}dp=\frac{e^{-m(t_2-t_1)}}{2m}
\end{equation}

The restriction $t_1\leq{t_2}$ has been done in order to make contact with the two point function. However, we will define $S_2(t_1;t_2)$ as $\frac{e^{-m(t_2-t_1)}}{2m}$ for any $t_1$ and $t_2$. (The $S$ stands for Schwinger).

\subsubsection{Gaussian process description and Feynman-Kac formula}
The {\it oscillator process} or {\it Ornstein-Uhlembeck process} is defined as a Gaussian process, indexed by $t\epsilon{R}$, with mean and covariance given by:

\begin{eqnarray}
E(\Phi_t)&=&0\\
E(\Phi_{t_1}\Phi_{t_2})&\equiv&{S}_2(t_1;t_2)
\end{eqnarray}

In order to prove that this process exist, we should verify that each $n\times{n}$ matrix $M$ of coefficients $M_{ij}\equiv{S}_2(t_i;t_j)$ (for an arbitrary choice of $t_i$, $i=1,...n$, with $t_i\neq{t}_j$) is positive-definite. That could seem a difficult exercise if we use the expression
$\frac{1}{m}e^{-(t_2-t_1)}$. However, using the integral representation $\int\frac{e^{ip(t_2-t_1)}}{p^2+m^2}dp$ it becomes clear that is a positive-definite matrix.

By combining the imaginary time extension of the two-point function with the Gaussian process description, we get the following formula:

\begin{equation}
(\Omega, \hat{\Phi}_0e^{-H_0(t_2-t_1)}\hat{\Phi}_0\Omega)=E(\Phi_{t_1}\Phi_{t_2})
\end{equation}

which is a particular and trivial case of the so-called {\it Feynman-Kac} formula. In the approach we have followed that formula is trivial because we have constructed
the Gaussian process by imposing that the l.h.s gives its covariance matrix. This formula can be generalized as follows:
{\center\underline{Feynman-Kac formula}}
\begin{equation}
\boxed{
(\Omega, \hat{\Phi}_0e^{-H_0(t_2-t_1)}..\hat{\Phi}_0e^{-H_0(t_n-t_{n-1})}\hat{\Phi}_0\Omega)=E(\Phi_{t_1}\Phi_{t_2}...\Phi_{t_n})}\label{FK}
\end{equation}

with $t_i\leq{t}_{i+1}$.

It can be obtained a further generalized expression, by replacing in this formula each insertion of $\hat{\Phi}_0$ in the position $i$ by any polynomial $F_i$.

\subsubsection{Feynman-Kac formula in the non-trivial case: path integral representation for an interaction}

As we have said, the expectations of products of Gaussian processes admit a functional integral representation. This corresponds to an integral over the field (depending on time) configuration. However, that
functional representation is unnecessary when it is computed the expectation of functions depending on Gaussian variables correspondent to a finite set of instants. Such is the case of the r.h.s of Eq. \ref{FK}. For this simple case, the integral is reduced to a multidimensional ordinary integral of a Gaussian type, having the measure of Eq. \ref{medidagaussiana}.

The functional integral representation is more useful when we consider vacuum correlation functions of the type $(\Omega, \hat{\Phi}_0e^{-Ht}\hat{\Phi}_0\Omega)$, with $H=H_0+ V$, being $V$  an operator describing a potential added to the harmonic oscillator, which is function (denoted also as $V$) of the position operator (with suitable conditions which we will not consider in this note).

In that case, it holds a non-trivial version of the Feynman-Kac formula, which takes the following form:

\begin{equation}
(\Omega,\hat{\Phi}_0e^{-H(t_2-t_1)}\hat{\Phi}_0\Omega)=E(\Phi_{t_1}\Phi_{t_2}e^{-\int_{t_1}^{t_2}V(\Phi(s))ds})
\end{equation}

for $t_1\leq{t_2}$.

In the r.h.s we see the expectation of factors including the function  $e^{-\int_{t_1}^{t_2}V(\Phi_s)ds}$ depending on a infinite set of Gaussian variables $\Phi_s$ for any real value $s$ in the interval $[t_1,t_2]$. An expectation of such type of function ({\it non-cylindric} according with the usual terminology) can not be written as a finite dimensional integral. If we insist in writing the r.h.s as an integral, we are forced to use a true functional integral over all the paths in the range $[t_1,t_2]$.

The previous formula can be generalized to the n-point functions as follows:

\underline{Non-trivial Feynman-Kac formula}

\begin{equation}
\boxed{(\Omega, \hat{\Phi}_0e^{-H(t_2-t_1)}...\hat{\Phi}_0e^{-H(t_n-t_{n-1})}\hat{\Phi}_0\Omega)=E(\Phi_{t_1}...\Phi_{t_n}e^{-\int_{t_1}^{t_n}V(\Phi_t)})}\label{FKinteracting}
\end{equation}

This formula admits a further generalization: we can replace each insertion of $\hat{\Phi}_0$ in the position $i$ by any polynomial $F_i$ (in particular the constant function).

\subsubsection*{Remark on the link with the Schrodinger representation of path integral}

In Eq. \ref{FKinteracting} the functional integration is taking over the set of all paths without any restriction. We are more familiar with a slightly different version of the previous path integral. It arises when it is computed the formal expression $<q''\mid{e}^{-(t_2-t_1)H}\mid{q'}>$, being $q>$ the eigenstates of the position operator with eigenvalues $q'$ and $q''$. In that case, the inner product can be written as path integral over the set of paths starting at $q'$ at $t_1$ and finishing at $q''$ in $t_2$:

\begin{equation}
<q''\mid{e}^{-(t_2-t_1)H}\mid{q}'>=\int e^{\int_{t_1}^{t_2}V(q(s))}DW^{(q',t_1;q'';t_2)}
\end{equation}

Here $DW^{(q',t_1,q'';t_2)}$ is the {\it conditional} measure associated to the oscillator process, arising by the restriction to those paths starting at $q'$ in $t_1$ and finishing at $q''$ at the instant $t_2$. If we want to use this measure for the vacuum
correlation function, we will get a more complicated expression, because the vacuum itself is (in the Schroedinger representation) the function $R$ given by: $R(x)= -\frac{1}{4}e^{-\frac{x^2}{2}}$ and the operator $\Phi_0$ is the multiplication by $x$. Taking this
into account, we can get the path integral representation:

\begin{eqnarray}
(\Omega,\hat{\Phi}_0e^{-H(t_2-t_1)}\hat{\Phi}_0\Omega)=\\\nonumber
\pi^{-\frac{1}{4}}\int\int{e}^{-\frac{q'^2+q''^2}{2}}q'q''\int e^{-\int_{t_1}^{t_2}V(q(s))ds}DW^{(q',t_1:q'',t_2)}dq'dq''
\end{eqnarray}

This expression is not useful for our purpose because in the QFT case, we will not use the coordinate representation (or Schrodinger representation) but the Fock representation.

\subsection{ The Gell-Mann-Low formula: the link between free and interacting vacuum correlation function:}

This is the most important formula because, according with the strategy mentioned in the introduction, it will be used for the definition of n-point functions in the interacting QFT case. In this formula, the vacuum and the Heisenberg position operators are those of the interacting theory. We will assume that the Hamiltonian $H$ is bounded from below and that it has a unique eigenvector $\Omega^{int}$ -the interacting vacuum- corresponding to the lowest eigenvalue $E$ of $H$.

\subsubsection{Interacting vacuum in terms of free vacuum}

 Under the previous assumptions (with further technical requirements) it can be derived the following relation:

\begin{equation}
\boxed{\Omega^{int}=\lim_{T\rightarrow\infty}\frac{e^{-TH}\Omega}{\sqrt{(e^{-TH}\Omega,e^{-TH}\Omega)}}}
\end{equation}

Such relation follows by expanding the free vacuum in terms of the eigenvalues of the interacting Hamiltonian. For later purposes we also include this useful formula, derived along the same line:

\begin{eqnarray}
E=-\lim_{T\rightarrow\infty}\frac{log(\Omega,e^{-TH}\Omega)}{T}=\\\nonumber
-\lim_{T\rightarrow\infty}\frac{log(E(e^{-\int_{0}^TV(\Phi_s)ds}))}{T}
\end{eqnarray}

which expresses the shift in the vacuum energy due to the interacting term as an expectation value in the free theory.
\subsubsection{Feynman-Kac plus Gell-Mann-Low}
As we have said, the central objects for the construction of the interacting quantum field theory are the interacting vacuum ($\Omega^{int}$) correlation functions of the interacting field $\hat{\Phi}^{int}_{t}$:

\begin{equation}
(\Omega^{int}, \hat{\Phi}^{int}_{t_1}...\hat{\Phi}^{int}_{t_n}\Omega^{int})\nonumber
\end{equation}

Actually, the useful quantity for the Gaussian process interpretation is the imaginary time extension of this quantity for $t_{i+1}\geq{t_i}$, by formally changing $t_j$ by $it_j$.

Using the previous relation between free and interacting vacuum (without being worried about commutation of the limits), we can get the following formula:

\begin{eqnarray}
(\Omega^{int}, \hat{\Phi}^{int}_{it_1}...\hat{\Phi}^{int}_{it_n}\Omega^{int})=\\\nonumber
\lim_{T\rightarrow\infty}\frac{(\Omega,e^{-H(T+t_1)}\hat{\Phi}_0{e}^{-H(t_2-t_1)}...e^{-H(t_n-t_{n-1})}\hat{\Phi}_0e^{-H(T-t_n)}\Omega)}
{(e^{-TH}\Omega,e^{-TH}\Omega)}
\end{eqnarray}

We have expressed the vacuum correlation function of the interacting quantum field in terms of a free correlation function of time zero field combined with exponential of the full Hamiltonian, which is of the form of the l.h.s of Eq. \ref{FK}.

By using a generalized form of Eq.[\ref{FKinteracting}] we get the following formula:
\bigskip

\underline{{\bf Euclidean Gell-Mann-Low formula}}

\begin{equation}
\boxed{(\Omega^{int}, \hat{\Phi}^{int}_{it_1}..\hat{\Phi}^{int}_{it_n}\Omega^{int})=
\lim_{T\rightarrow\infty}\frac{E(\Phi_{t_1}...\Phi_{t_n}e^{-\int_{-T}^{T}V(\Phi_s)ds})}{E(e^{-\int_{-T}^{T}V(\Phi_s)ds})}}\label{EGL}
\end{equation}

for $t_{j+1}\geq{t_j}$.

\section{Difficulties arising in $D=1+1$}

The addition of one dimension force us to treat the field operator as a {\it distribution}. That is the source of difficulties for one of the steps of the construction of the interacting QFT model. There is not a problem in the Gaussian process description; in fact, this is an straightforward generalization of the harmonic oscillator case, by a suitable replacement of the covariance. However, as we have anticipated, the distributional character
of the field is the source of the difficulties for the definition of the interaction term like $\lambda:\Phi^4:$

Let us go with the first part of the construction: the Gaussian process description of the free field.

\subsection{Covariance and the two point function of a free scalar field}

The required Gaussian processes are now indexed by a {\it function}. More precisely, the index will change from the real value $t$ to the pair $t,h$, being $h$ a function of the spatial coordinate. The covariance of the associated Gaussian process will be defined in terms of the 2-point function of the field.

Let be $f$ and $g$ functions of the spatial coordinate belonging to the Schwartz space $S(R)$. We will not explain the motivation for this technical condition. The only important thing for us is that the functions $f$ and $g$ should vanish at infinity. Then, these functions are not allowed to be constants. We define the following vacuum correlation:
\begin{equation}
W_2(t_1,h_1;t_2,h_2)\equiv{<}0\shortmid\hat{\Phi}_{t_1}(h_1)\hat{\Phi_{t_2}}(h_2)\shortmid0>
\end{equation}

As we have said in the introduction, $W_2$ is a distribution that can not be written as an integral of the form:
$\int\int W_2(t_1,x_1;t_2,x_2)h_1(x_1)h_2(x_2)dx_1dx_2$. However, it is useful such formal expression. Having this abuse of language in mind, we consider $W_2$ as a function of the spacetime points. It is known that
this 2-points function can be written as:

\begin{equation}
W_2(t_1,x_1;t_2,x_2)=\int\frac{e^{-ik(x_2-x_1)+i\sqrt{k^2+m^2}(t_2-t_1)}}{\sqrt{k^2+m^2}}dk
\end{equation}

Although the r.h.s is not well defined everywhere, the previous equation is a distributional statement which has a precise meaning.

Using the relation $\frac{1}{2\pi}\int\frac{e^{ip(t_2-t_1)}}{p^2+m^2}dp=\frac{e^{-m(t_2-t_1)}}{2m}$, we can show that the analytic continuation $W_2(it_1,x_1;it_2,x_2)$ to imaginary time for $t_2>t_1$ is:

\begin{equation}
W_2(it_1,x_1;it_2,x_2)=\int\frac{e^{ik.(x_2-x_1)+ip(t_2-t_1)}}{k^2+p^2+m^2}{d}kdp
\end{equation}

The left hand side can be extended to any pair $t_1$ and $t_2$ and it will define the 2-points Schwinger function:\\

\underline{Schwinger 2-points function of D=2 scalar field}
\begin{equation}
\boxed{
S_2(t_1,x_1;t_2,x_2)\equiv\int\frac{e^{ik(x_2-x_1) + ip(t_2-t_1)}}{k^2+p^2+m^2}{d}kdp}\label{covarianceD2}
\end{equation}

As in the previous case, if we want to avoid abuse of language, we should consider $S_2$ a distribution which need two spatial functions as entries:

\begin{equation}
S_2(t_1,h_1;t_2,h_2)\equiv\int\frac{\overline{\hat{h}_1(k)}e^{ip(t_2-t_1)}h_2(k)}{k^2+p^2+m^2}{d}kdp
\end{equation}

The r.h.s of the previous equation is well defined for any pair $t_1,t_2$, without the restriction $t_2\geq{t}_1$. Notice that this quantity is symmetric in its argument, which is consistent with its interpretation as the expectation of commutating fields.

Because $S_2(.,.;., .)$ is a positive-definite matrix in $R\times{S}(R)$, we know by the Kolomogorov theorem that it will exist a Gaussian process $\Phi_{t,h}$, indexed by the pair $t,h$ (being  $h$ a function belonging to the Schwartz space), having zero mean and covariance given by $S_2(..;.. )$. Moreover, it can be shown that $\Phi_{t,.}$ is a {\it linear functional} on on $S(R)$. Then, the Gaussian process inherit the distributional character of the operator value distribution which was used for the definition of the covariance.

\subsection{The free Feynman-Kac formula}

 The previous formulas for the case of the harmonic oscillator have an straightforward generalization by replacing the index $t$ by a $t,h$. The free Feynman-Kac generalized formula reads:

\begin{equation}
(\Omega,\hat{\Phi}_0(h_1)e^{-H_0(t_2-t_1))}...\hat{\Phi}_0(h_n)e^{-H_0(t_n-t_{n-1})}\Omega)=E(\Phi_{t_1}(h_1)\Phi_{t_2}(h_2)...\Phi_{t_n}(h_n))
\end{equation}

with the condition $t_n\geq{t}_{n-1}$.

\subsection{A digression: Gaussian processes indexed by a spacetime function}

For further purpose it could be convenient to threat on equal foot space and time by considering a Gaussian process $\Phi(f)$ indexed by a function $f$ of the spacetime, such that its covariance takes this form:

\begin{equation}
E(\Phi(f_1)\Phi(f_2))=\int\frac{\overline{\hat{f}_1(k,p)}f_2(k,p)}{k^2+p^2+m^2}{d}kdp
\end{equation}

The relation between both descriptions is given by:

\begin{equation}
\Phi(f)=\int \Phi_t (f_t) dt
\end{equation}

being $f_t$ the function of a single variable $x$ such that $f_t(x)=f(t,x)$. We can be convinced of it formally by written each side as an integral of $\Phi(x,t)$ and $\Phi_t(x)$ smeared with $f(x,t)$ and $f_t(x)$ respectively.

\subsection{Wick products as polynomials}

In order to define interactions, we will need also the so-called {\it Wick products} among Gaussian variables. For monomial expressions of
degree $n$, these are defined as polynomials of degree $n$, including lower powers terms:

\begin{equation}
:\Phi(h_1)\Phi(h_2)...\Phi(h_n):\equiv\Phi(h_1)\Phi(h_2)...\Phi(h_n)+ lower\; order\; terms.
\end{equation}

The coefficients of the lower order terms are fixed by the following conditions which define the Wick products:

\begin{itemize}
\item $:\Phi(h):=\Phi(h)$
\item $E(:\Phi(h_1)\Phi(h_2)...\Phi(h_n):)=0$
\item $E(:\Phi(h_1)\Phi(h_2)...\Phi(h_n)::\Phi(h_1)\Phi(h_2)...\Phi(h_m):)=0$ for $n\neq{m}$
\end{itemize}

For instance,

\begin{equation}
:\Phi(h_1)\Phi(h_2):=\Phi(h_1)\Phi(h_2)-E(\Phi(h_1)\Phi(h_2))
\end{equation}

In the general case, the coefficients of the polynomials will be combinations of the two point functions $E(\Phi(h_i)\Phi(h_j))$.

An important case for us is the definition of $:(\Phi(h))^4:$:

\begin{equation}
:(\Phi(h))^4:=(\Phi(h))^4-6.(S_2(h,h))(\Phi(h))^2 + 3 (S_2(h,h))^2
\end{equation}

As we have said, if we replace the Gaussian process by the operator $\hat{\Phi}(h)$ (and the Schwinger function $S_2$ replaced by the time ordered 2-point functions $W_2$) we will get the usual Wick ordered expression, arising after moving the creator operator to the left. The definition as a polynomial will be useful for the control of divergences when we define the interacting term.

\subsection{Reflection Positivity (RP)}

The Schwinger functions inherit two types of positivity conditions: one coming from its interpretation as expectation of Gaussian processes, which can be always written as a positive defined inner product. The other positivity condition comes from their definition as the imaginary time extension of the n-point functions, which are also coming from an inner product in the Hilbert space. The last positivity condition is translated in a set of inequalities involving Schwinger functions. This set of conditions is known as {\it reflection positivity}(RP). As we have mentioned, it is one of the OS axioms \cite{OS}.

We will not reproduce the proof of RP starting from W axioms  but we will simply show how RP is derived in the simple case of an scalar free field. Let us consider the vector: $e^{-Ht}\hat{\Phi}_0(h)\Omega$, for $t\geq0$. By taking the norm of it and rewriting the norm in terms of Schwinger functions, we will find the simplest case of the inequalities, which involves only Schwinger 2-points functions:

\begin{equation}
\parallel{e^{-Ht}\hat{\Phi}_0(h)\Omega}\parallel^2=(\Omega,\hat{\Phi}_0(h)e^{-Ht}e^{-Ht}\hat{\Phi}_0(h)\Omega)
=S(-t,h;t,h)\geq{0}
\end{equation}

 That is the simplest case of the set of inequalities encoded in RP. A more complicated inequality arise if we define the following vector:

\begin{equation}
v\equiv{e}^{-Ht_1}\hat{\Phi}_0(h_1)\Omega+ {e}^{-Ht_2}\hat{\Phi}_0(h_2)\Omega + e^{-Ht_1}\hat{\Phi}_0(h_1)e^{-H(t_2-t_1)}\hat{\Phi}_0(h_2)\Omega
\end{equation}

for $t_2\geqq{t}_2$ and we take the norm:
\begin{eqnarray}
0\leqq\parallel{v}\parallel^2=S(t_1,h_1;-t_1,h_1)+S(-t_2,h_2;-t_2,h_2)+\\\nonumber
S(-t_2,h_2;t_1,h_1)+S(-t_1,h_1;t_2,h_2)+S(-t_1,h_1;t_1,h_1;t_2,h_2)\\\nonumber
 + S(-t_2,h_2;t_1,h_1;t_2,h_2)+S(-t_2,h_2;-t_1,h_1;t_1,h_1;t_2,h_2)
\end{eqnarray}

As we can see, the inequalities will involve more an more terms, with an increasing number of Schwinger functions, some of them containing the change in the sign of the time index.

\subsubsection*{General statement of RP}

 RP can be rewritten in a form adapted to the general case in which we can not assume the existence of fixed time field. In that case we will need a spacetime test function. The restriction to positive instant $t_i$ in the fixed time free field will be translated in the restriction of the support of the test function to the upper plane $R\times[0,+\infty)$. The change in the sign of $t_i$ occurring in the previous inequalities will be translated in the the application of an operation $\Theta$ acting on a spacetime functions as: $\Theta(f)(x,t)= f(x,-t)$.

In order to give a more precise statement of RP, let us introduce a family of functions $\{f_j\}$, having supports included in $R\times[0,+\infty]$ and ``chronologically ordered'', i.e., the instants in which $f_i$ is not vanishing should be less or equal than the instants in which $f_{i+1}$ is not vanishing. Let us introduce also the following notation:
$A^{(n)}_{f}$ for a sum of product of the field on the form: $\Phi(f_1) + \Phi(f_2) + ..\Phi(f_1)\Phi(f_2)..\Phi(f_n)$ (arising from making all the combination of products of $\Phi(f_i)$ for $i=1..n$ up to $n$ factors. RP
can be written as the following statement:

\begin{equation}
E(\theta(A^{(n)}_f),A^{(n)}_f))\geq0\label{RPutil}
\end{equation}

This form of RP is more useful for the case of interacting case arising by a perturbation of a Gaussian measure. We will use such expression later.
\newpage
\part{Constructing $\lambda\Phi^4$ in $D=1+1$}

\newpage
\section{Facing the first divergence: defining the cut-off interacting term}\label{dospuntosSection}

\subsection{The difficulties for defining powers of a Gaussian field}

In order to define an interaction term, we would like to define a power of the Gaussian variable as a new random variable, which should be a linear functional. We are not referring to expressions like $(\Phi(.))^4$ which takes a function $h$ and gives as an output $(\Phi(h))^4$. That is not what we are looking for
because it is not a linear functional.

As we have mentioned at the beginning, we can consider a family of functions $h_{\kappa}^{(x)}$, labeled by an integer number $\kappa$,
localized around $x$, such that as $\kappa\rightarrow\infty$, $h_{\kappa}^{(x)}$ approach to the $\delta_x$ in a distributional sense (for instance, we can consider a family of functions of the form $h_{\kappa}^{(y)}(y)=\frac{sin(\kappa(x-y))}{x-y}$).

Then, we can consider an operator valued regular distribution
$\Phi_{\kappa}^n$, indexed by $\kappa$, whose kernel is defined by $\Phi_{\kappa}^n(x)\equiv{({\Phi}(h_{\kappa}^{(x)}))}^n$. We should take into account that here $\Phi_{\kappa}^n(x)$ is defined as the kernel of distribution $\Phi_{\kappa}^n$ that we want to define. That means that its action on a function $f$ is defined by:

\begin{equation}
\Phi_{\kappa}^n(f)=\int\Phi_{\kappa}^4(x) f(x) dx.
\end{equation}

However, the existence of the limit $\kappa\rightarrow\infty$ for the functional ${\Phi_{\kappa}}^n$ is not guarantied . This is a particular case of the usual problems in defining a product of distributions. In the QFT context that is usually rephrased as the divergence problem appearing in the coinciding point limit of products of fields in different spacetime points.

This divergence could be avoided if we take the limit $\kappa\rightarrow\infty$ in the Wick power of the field. That is:

\begin{equation}
:\Phi^n:(x)\equiv\lim_{\kappa\rightarrow\infty}:{\Phi(h_{\kappa}^{(x)})}^4:
\end{equation}

 Due to the some features of the Wick product $::$, there is more chance for the existence of the limit. As usual, here we are making the abuse of language in writing  $:\Phi^n:(x)$ instead of the more appropriated $:\Phi^n:(h)$.

The existence of that limit depends crucially on the power $n$ and also on the dimension of the spacetime. For instance, for $n=2$ this limit exists in any dimension \footnote{In fact, as we have said before, the previous procedure for the definition of the expression $:\Phi^n:(x)$ has its counterpart in the operator approach as the normal order procedure. And we know that we can always define (rigorously) the number operator as the normal ordered of square of the field. In a similar way, we can define the free Hamiltonian as the normal ordered version of quadratic combinations of the field operator. We will go back to this point later}. However, for a bigger power, there is not guaranty of its existence. Let us consider the relevant case $n=4$

\subsection{In $D=2$ it can be defined a cut-off interacting term}

In order to define $:\Phi^4:(x)$, we need that the following limit exists:

\begin{equation}
\boxed{
:\Phi^4:(x)\equiv\lim_{\kappa\rightarrow\infty}(\Phi_{\kappa}(x))^4-6c_{\kappa}(\Phi_{\kappa}(x))^2+3c_{\kappa}^2}
\end{equation}
where $c_{\kappa}$ is the covariance $S_2(h_{\kappa}^{(x)},
h_{\kappa}^{(x)})$. We have omitted the dependence on $x$ in the coefficient $c_{\kappa}$ because we can see it becomes independent of $x$ if we make an appropriate choice of the functions $h_{\kappa}^{(x)}$  (like the one mentioned before).

Let us notice that in the limit $\kappa\rightarrow\infty$ the coefficient $c_{\kappa}$ diverges as $log(\kappa)$ in $D=2$. This behaviour follows by observing that in this limit the function $h_{\kappa}^{(x)}$, which is the argument of the covariance,  become a function sharply concentrated at $x$  and so $c_{\kappa}$ approaches to the formal integral: $\int\frac{1}{k^2+m^2}{d}^2k$. The fact that this has a logarithmic divergence could be checked by using known results about the singularity of the two point function in $D=1+1$ for the coinciding points limit. And due to its particular type of divergence, it can be proven that the previous limit exists. That is the kind of proof that we have decided to omit.

Having defined the distribution $:\Phi^4:$, we can define the interacting term by applying it to an space-time function $f$ vanishing at infinity. In particular, we can take this function as $\lambda\chi_{\Lambda}$, being $\chi_{\Lambda}$ the characteristic function on a bounded spacetime region $\Lambda$. So, the interaction term need for the perturbation of the Gaussian measure is:\\

\begin{equation}
\boxed{
\frak{A}^{\Lambda}\equiv\;:\Phi^4:(\lambda\chi_{\Lambda})=\int_{\Lambda}:\Phi^4:(x) d^2x}
\end{equation}

\subsubsection{The justification of formal manipulation}

From the definition of $:\Phi^4:$ and the properties of the Wick products, we can show that the following useful formula holds:

\begin{equation}
E(:\Phi(h_1)\Phi(h_2)\Phi(h_3)\Phi(h_4)::\Phi^4:(g))=4!\int[\prod_{j=1}^4(h(x_j)S_2(x_j,y)]g(y)dxdy
\end{equation}

 This formula can also be derived by a formal manipulation, considering the expression $:\Phi^4:(x)$ as an ordinary Wick product of the form $:\Phi(x_1)\Phi(x_2)\Phi(x_3)\Phi(x_4):$ in the limit when all the $x_i$'s approaches $x$ and using the usual Wick contraction theorem. So, even though $:\Phi^4:$ has not a simple expression in terms of the original Gaussian variables, it can be manipulated very easily. Part of the technicalities of CQFT are related with the derivation of useful bounds for expectations involving this object which appears in the interacting term.

\section{Dealing with a second divergence: the quantum boundedness from below}\label{StabilitySection}

Before taking the infinite volume limit, we should verify that the exponential of the cut-off interaction is integrable. That is necessary condition for the finiteness of the n-point function. As we have mentioned, although the polynomial $F$ given by $F(x)=\lambda{x}^4$ is bounded from below as a real function, the interacting term is
defined as $\lambda\int_{\Lambda}:\Phi^4:(x)$. The Wick product $:\;:$ destroys the positivity of the operator, as we can see from the definition:

\begin{equation}
:\Phi^4:(x)=\lim_{\kappa\rightarrow\infty}P_{\kappa}(\Phi_{\kappa}(x))\nonumber
\end{equation}

being $P_{\kappa}$ a polynomial of 4-degree of the form: $P_{\kappa}(z)=z^4-6c_{\kappa}z^2+3(c_{\kappa})^2$. This lowest value of $P_{\kappa}$ is $-6{c_{\kappa}}^2$. Taking into account the behavior of $c_{\kappa}$ for large value of $\kappa$, we see that the deep of this minimum goes as $-(log(\kappa))^2$ (See Fig. 1)

\begin{figure}[t]
  \centering
  \includegraphics[width=60mm]{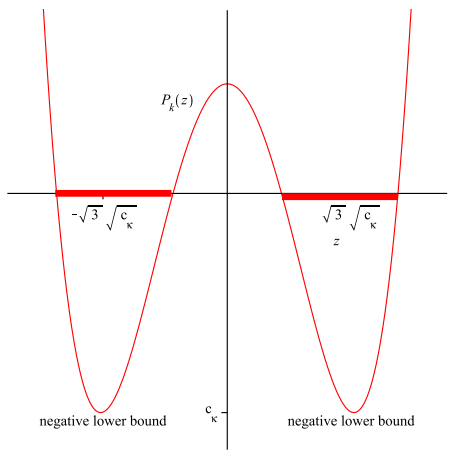}\\
  \caption{Due to the Wick product, the $\kappa$ dependent interacting term becomes a polynomial $P_{\kappa}$ which takes negative values. Its lower value is $-c_{\kappa}$, which becomes arbitrarily negative as $\kappa\rightarrow\infty$}\label{LowerFig}
\end{figure}

The relevant information for the convergence of the integral $\int e^{-\mathfrak{A}^{\Lambda}} d\mu$  is the size of the region in the field configuration space in which $:\Phi^4:(x)$ takes this negatives values. As we will see, this size is small enough for the convergence of the integral.

\subsection{A preliminary observation: powers of the interaction term are integrable}
Before considering the proof of the integrability of the exponential term, we should explain why the following naive argument does not work: if we make a formal Taylor expansion of the exponential (as function of $\lambda$ around $\lambda=0$) inside the integral $\int e^{-\frak{A}^{\Lambda}} d\mu$ and distribute the integral, we will  get a series whose generic term is an integral of powers of $\frak{A}^{\Lambda}$. It can be shown that
$\int {(\frak{A}^{\Lambda})}^n d\mu<\infty$. Proving the integrability of each term of the Taylor expansion seem to be enough for the proof of the integrability of the exponential.

Here is where the issue of the non-convergence of the series enters. If the formal Taylor series were (fast enough) convergent to the exponential, we could use the previous result in order to prove easily the integrability of the exponential. However, the divergent character of the series does not allow this kind of proof. This subtle is not very surprising because even though we denote these expressions by the name `exponential' and `power', the nature of the space in which this expression are integrated makes the issue more complicated than in the case of an ordinary single variable.

\subsection{The proof of stability}

We have to proof that $E(e^{-\frak{A}^{\Lambda}})=\int e^{-\frak{A}^{\Lambda}}d\mu <\infty$. It could seem a difficult task because this is a functional integral. However, we could rewrite this integral as an ordinary Lebesgue integral:
\begin{equation}
\int e^{-\mathfrak{A}^{\Lambda}}d\mu=\int_{0}^{+\infty}h(t) dt\label{integral}
\end{equation}

being $h(t)=\mu\{\Phi: e^{-\mathfrak{A}^{\Lambda}}>t\}$. So, what we need to prove is that the function $h$ decrease fast enough to make the integral convergent. Because we
are afraid of a divergence when $\kappa\rightarrow\infty$, it is enough to see the behavior of the function $h$ for large $t$.

Here we should appeal to a technical result on some Gaussian integrals which appear in most of the standard exposition (\cite{Jaffe1}). There are two key inequalities:

\begin{eqnarray}
\mathfrak{A}^{\Lambda}_{\kappa}>-N(log\kappa)^2\label{Bound1}\\
\int(\mathfrak{A}^{\Lambda}-\mathfrak{A}^{\Lambda}_{\kappa})^2\leq\alpha{e}^{-\beta{\kappa}^{1/4}}\label{Bound2}
\end{eqnarray}

for $N,\alpha,\beta$ are positive constants independent of $\kappa$. The first inequality follows from the lower bound we have mentioned; the second characterize the precise speed of the convergence of the approximate interacting term $\mathfrak{A}_{\kappa}$ to the $\mathfrak{A}$.

Let us choose a large value of $t$ in the way: $t_{\kappa}=e^{N(log\kappa)^2-1}$ in Eq.[\ref{integral}]. As far as $\kappa$ goes to $\infty$ this $t_{\kappa}$ covers all the real values of $t$ from certain positive value on. From the previous inequalities it follows a bound for large value of $t$ in the function $h$:

\begin{equation}
h(t=e^{N(log\kappa)^2-1})=\mu\{\Phi: \mathfrak{A}^{\Lambda}<-N(log\kappa)^2+1\}\leq\mu\{\Phi: \mathfrak{A}^{\Lambda}- \mathfrak{A}^{\Lambda}_{\kappa}<-1\}
\end{equation}

where in the last step we have used (\ref{Bound1}). Using now (\ref{Bound2}) , we get that for large value of $t$:

\begin{equation}
h(t)< \mu\{\Phi: |\delta\mathfrak{A}^{\Lambda}|>1\}\leq\int(\delta\mathfrak{A}^{\Lambda}_{\kappa})^2\leq\alpha{e}^{-{exp}(1/4\sqrt{\frac{log(t)+1}{N}})}
\end{equation}

So, $h$ is a positive function bounded by a integrable function. Then, we have proved the convergence of the wished integral.

\section{Dealing with the last risk of divergence: the infinite volume limit}\label{volumeninfinito}

We are now in the more difficult part. We still have to prove that the previous cut-off Schwinger functions converge when $\Lambda\rightarrow\infty$. In this step, we will be still more schematic than we were before.

We start mentioning that the reason why the limit $\Lambda\rightarrow\infty$ could give rise to a divergence is related with the low decay of the covariance -i.e., the two point function $S_2$- used for the definition of the Gaussian processes that we have considered. In order to see this relation in a heuristic way, it could be useful to consider modified Schwinger functions arising after replacing the standard Gaussian processes with new ones defined by a modified covariance fulfilling a wide class of Dirichlet conditions. This Dirichlet conditions eliminate the low decay behaviour of the covariance. After observing in the next subsection that the issue of the convergence of the Schwinger functions becomes trivial in these cases, we will see how the so-called {\it cluster expansion} makes a clever use of this trivial fact in order to prove that the infinite volume limit exists in the case in which the standard covariance $S_2$ is used.

\subsection{Imposing Dirichlet conditions makes trivial the problem of convergence}

\subsubsection{Dirichlet conditions on the covariance}

Without entering in precise definitions, we want to mention that it is possible to define univocally a family of modified covariances $C^{\Gamma}(x,y)$, fulfilling the condition $C^{\Gamma}(x,y)=0$ for $x$ or $y$ belonging to a certain path $\Gamma$ in the $R^2$. Any member of this family is defined by the inverse of the operator $-\Delta_{\Gamma} + m^2$, being $\Delta_{\Gamma}$ the Laplacian operator acting on the subset of functions of $L_2$ vanishing in $\Gamma$. The usual covariance $S_2$ we have used until now can be obtained by the inverse of  $-\Delta + m^2$, where $\Delta$ is the standard Laplacian acting in the whole space of functions of $L_2$, free of any Dirichlet condition. Because of that, we will call it {\it free covariance}

In particular, it will be relevant the case in which $\Gamma$ is any finite union of the unit segments which are the boundary of the lattice unit squares of $R^2$ (the dotted lines in Fig. 2). This family of covariances includes two extremal cases:

\begin{itemize}
\item $\Gamma=\emptyset$. It corresponds to the  free covariance.
\item  $\Gamma=B$, being $B$ the entire grid displayed in Fig. 2. It corresponds to the {\it completely decoupled} covariance.
\end{itemize}

In the second case $C^B(x,y)=0$ for $x$ and $y$ belonging to different unit squares. That justifies the name ``decoupled". What we want to emphasize is that any member of this family of covariances is suitable for the definition of Gaussian processes because it defines a positive-definite inner product. Even more, this family of covariances fulfills reflection positivity. That is a property that we want to keep.

For the next considerations, we will need a bigger family of covariances interpolating between different members of the previous  discrete family. Let us consider for instance a unit segment $\Gamma_0$. We want to find a continuous family of covariances $C^s$ (with $s$ in $[0,1]$) which interpolates between $C_{B-\Gamma_0}$ and $C_{B}$. One possibility could be:
\begin{equation}
C^{s}\equiv{(1-s)}C_{B} + sC_{B-\Gamma_0}
\end{equation}

The role of the parameter $s$ is to assign a weight to the Dirichlet condition on the segment $\Gamma_0$. That means that for $y\epsilon\Gamma_0$ and a generic value of $s$,  $C^s(x,y)$ will be different from zero; for $s=1$, we will have non Dirichlet condition on $\Gamma_0$ and $s=0$ correspond to the case in which we have the full Dirichlet condition on $\Gamma_0$.

The previous linear combination can be generalized. In order to do that, we need to introduce an infinite components vector $s=(s_1,s_2,......)$ whose entrances introduce a weight for the unit segment $b_i$ in the entire grid $B$. In the main reference \cite{Jaffe3} we can find the following definition for the interpolating covariance:

\begin{equation}
C^{s}=\sum_{\Gamma}\prod_{j/b_j\epsilon{\Gamma}}s_j\prod_{i/b_i\epsilon{B}-\Gamma}(1-s_i)C_{B-\Gamma}
\end{equation}
where the sum over $\Gamma$ includes the vacuum set $\emptyset$.

The previous expression is not important for the next discussion. What we want to remark is just the existence of a vector allowing a continuous transition between the members of the family $\{C_{\Gamma}\}$. In particular:

\begin{itemize}
\item $s=(1,1,1,....)$ correspond to the free covariance $C_{\emptyset}$
\item $s=(0,0,0,....)$ correspond to the full decoupled covariance $C_{B}$.
\end{itemize}

\subsubsection{The infinite volume limit in the case of fully decoupled measure}
Let us consider now the Schwinger function $S_B^{(\Lambda)}(x_1,x_2,...,x_n)$ defined as the expectation value
$\frac{\int\Phi(x_1)...\Phi(x_n)e^{-\frak{A}^{\Lambda}}d\mu^{B}}{\int e^{-\frak{A}^{\Lambda}}d\mu^{B}}$ , in which the free Gaussian measure has been replaced by the completely decoupled measure $d\mu^B$. For convenience, the coefficient $c_{\kappa}$ used for the definition of the Wick product in $\frak{A}^{\Lambda}$ is still defined in terms of the free covariance as $S_2(h_{\kappa}^{(x)},h_{\kappa}^{(x)})$. That is the reason why this condition on the Schwinger functions is referred as {\it half Dirichlet} BC (see \cite{Simon}).

Let us consider the limit $\Lambda\rightarrow\infty$ of the Schwinger function $S^{(\Lambda)}_{B}(x_1,x_2,...,x_n)$, for points $x_1,x_2,...,x_n$ living in a bounded region $\Lambda_0\subset\Lambda$, consisting in the union of those unit squares containing at least one of the points $x_1...x_n$ (see Fig. 3). It is immediate to see that this Schwinger function converges in the limit $\Lambda\rightarrow\infty$:

\begin{eqnarray}
S_{B}^{(\Lambda)}(x_1,x_2,...,x_n)=\frac{\int \Phi(x_1)...\Phi(x_n)e^{-\frak{A}^{\Lambda}}d\mu^{B}}{\int e^{-\frak{A}^{\Lambda}}d\mu^{B} }=\\\nonumber
\frac{\int \Phi(x_1)...\Phi(x_n)e^{-\frak{A}^{\Lambda_0}}d\mu^{B}}{\int e^{-\frak{A}^{\Lambda_0}}d\mu^{B}}
\end{eqnarray}

Because the r.h.s of the last equation does not depend on $\Lambda$ but on $\Lambda_0$, it is clear that the limit
$\Lambda\rightarrow\infty S_{B}^{(\Lambda)}(x_1,x_2,...,x_n)$ exists.

\begin{figure}[t]
 \centering
 \includegraphics[width=50mm]{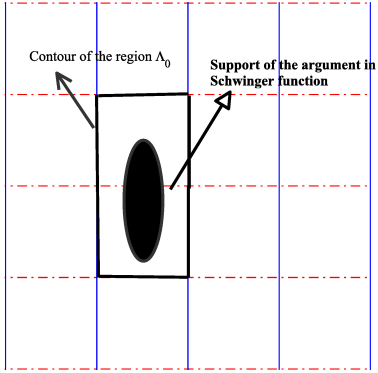}
\caption{Imposing Dirichlet conditions on the dotted lines makes trivial the problem convergence of the Schwinger functions. The black region denote the support of the functions which are arguments of the Schwinger functions. The region $\Lambda_0$ is the smallest set of unit squares containing the support region. Such region is independent of the cut-off region $\Lambda$}
\end{figure}

\subsection{A schematic oversimplified exposition of the cluster expansion}

Although this is an schematic exposition, we want to say more than: "after some hard computation it was showed that the infinite volume limit exists". We wish to give a feeling about this procedure because one of the non trivial steps of CQFT is the control of this divergence. In fact, only by taking a look at this step we can understand the increasing difficulties when we go to the case $D=2+1$ and $D=3+1$.

What makes the account of this step more complicated is the fact that there is not a single procedure for the proof of the existence of the infinite volume limit. We will consider here only one method: the so-called {\it cluster expansion}. It was applied in \cite{Spencer} to a general class of models describing interactions given by polynomials bounded from below. The family of these models, which include the one of this note, is called $P(\Phi)_2$. There are other tools (see \cite{Simon}) which work for the restricted family of polynomial of even degree plus a linear term, which also include our case $\lambda\Phi^4$. The cluster expansion and the proof of the convergence in the infinite volume limit involve several intermediate inequalities and a lot of definitions referring to different types of graphs. For a nice and precise account of this we recommend the reading of \cite{Spencer2} or Chapter 18 of the book \cite{Jaffe1}.

Besides the technical details, the ideas behind the cluster expansion are simple. One of these is the previous observation that {\it convergence becomes trivial when a Dirichlet condition is imposed on the boundary of $\Lambda$}. The cluster expansion make use of such observation by expressing the Schwinger functions with free BC as a series in which each term has a Dirichlet conditions in the entire lattice boundary lines $B$ with the exception of a finite
length path $\Gamma$ (a finite union of unit lattice segment), which labels the terms of the expansion. The goal is to have control of the infinite volume limit by expressing the free (=fully coupled) Schwinger functions in terms of almost decoupled quantities, which are under control.

The steps involved in the cluster expansion and its use for proof of the convergence in the infinite volume limit are the following:

\begin{enumerate}

\item {\bf The weak influence of far away boundary conditions}\\

First, it is proven that the Schwinger functions fulfill a property called  {\it regularity at infinity}. Let us explain what states this property in the simple case of the cut-off Schwinger n-point functions $S_{Free}^{\Lambda}$. As we have mentioned, we can change the free measure in several ways by using a Gaussian measure associated to a covariance $C_{\Gamma}$. Let considered the case in which $\Gamma=B-\Gamma_0$, being $\Gamma_0$ a finite union of lattice unit segments. The Schwinger function $S^{\lambda}_{B-\Gamma_0}$ with this modified measure will be a function of $\Gamma_0$.

In this case, regularity at infinity states that the following equality holds:

\begin{equation}
S_{Free}^{\Lambda}=\lim_{\Gamma_0\nearrow{B}}S_{B-\Gamma_0}^{\Lambda}
\end{equation}

where we have omitted the index $n$.

As the path $\Gamma_0$ increase, the Dirichlet conditions used for the definition of $S_{B-\Gamma_0}^{\Lambda}$ are confined in distant lattices segments in $B-\Gamma_0$. Hence, what states regularity at infinity is that these boundary conditions have a weak influence on the n-point functions $S_{\Gamma_0}^{\Lambda}(x_1,x_2,...,x_n)$ if $B-\Gamma_0$ is located in a far region. This weak influence vanishes in the limit $\Gamma_0\nearrow{B}$ in which we get the free boundary condition.

The previous property makes precise the notion of the weak influence of boundary conditions located far away. Although plausible, it should be proved. That is the less complicated part of the proof. That property is essential for the following steps, because allows to use Dirichlet covariances as a good approximation to the free covariances.

\item {\bf  Expressing free measure in terms of Dirichlet measure}

That is the second part of the cluster expansion, which is not very complicated. However, it involves an expansion which is not frequently used in physics. In order to introduce the idea behind this expansion, let us consider the following simple examples: If we take a function $f$ of a single variable, we can express the value $f(1)$ as follows: $f(1)= f(0)+ \int_0^1 f'(x) dx$. For a function of two variables, we can write:

\begin{equation}
f(1,1)=f(0,0) + \int_0^1\partial_xf(x,0) dx + \int_0^1 \partial_yf(0,y) dy + \int_0^1\int_0^1\partial_x\partial_y f(x,y) dxdy
\end{equation}

By a repeated use of the identity $\int_a^b f'(x) dx= f(b)-f(a)$, the previous expansion can be applied to a function of an arbitrary number of variables. The idea is to write $f(1,1,......1)$ as a sum of expressions containing $f$ and its partial derivatives evaluated in points with a decreasing numbers of $0$ in some of the coordinates. Of course, there is nothing special in the use of the point $(1,1,...)$ in the l.h.s; we can replace it for any other value, changing also the upper limit in the integral of the r.h.s.

Why this trivial identity could be useful for us? Let us recall that a Schwinger function corresponding to a generic covariance (belonging to the family we have mentioned) is a function of the infinities variables $s_i$, one of each giving a measure of the coupling  across a particular unit segment. The point $s=(1,1,....1)$ corresponds to the free boundary case, and the $0$'s in some entrances of $s$ says that there are Dirichlet conditions on the corresponding unit lattice segments. The goal is to express the Schwinger functions corresponding to $s=(1,1,...1)$ as a sum of quantities (derived from the Schwinger functions)  corresponding to other values of $s$ with many $0$ in their entrances.

In this step we see the importance of having a continuous range of values for each $s_i$: that allows to compute derivatives of the Schwinger functions with respect to these parameters and applied the previous expansion.

A minor remark: what it will be expanded is not the Schwinger function but the product: $Z^{\Lambda}S^{\Lambda}$, being $Z^{\Lambda}\equiv\int e^{-\mathfrak{A}^{\Lambda}}d\mu$ the partition function. The technical reason behind this choice is the following: the cluster expansion is an expansion of the free Gaussian measure $d\mu$ in terms of the others almost decoupled measures. Because the combination $Z^{\Lambda}S^{\Lambda}$, rather than $S^{\Lambda}$, is an expression of the type $\int....d\mu$, it is more natural to apply the previous expansion to $Z^{\Lambda}S^{\Lambda}$ rather than $S^{\Lambda}$.

We can illustrate this expansion by considering an already almost decoupled quantity. Let us consider, for instance, the Schwinger function correspondent to Dirichlet conditions in all the lattice grid $B$, with the exception of a unit square $\Box$. If we denote this Schwinger function as $S_{(B-\Box)}$ (omitting the spacetime n-point), the previous expansion takes the following form:

\begin{equation}
\includegraphics[width=50mm]{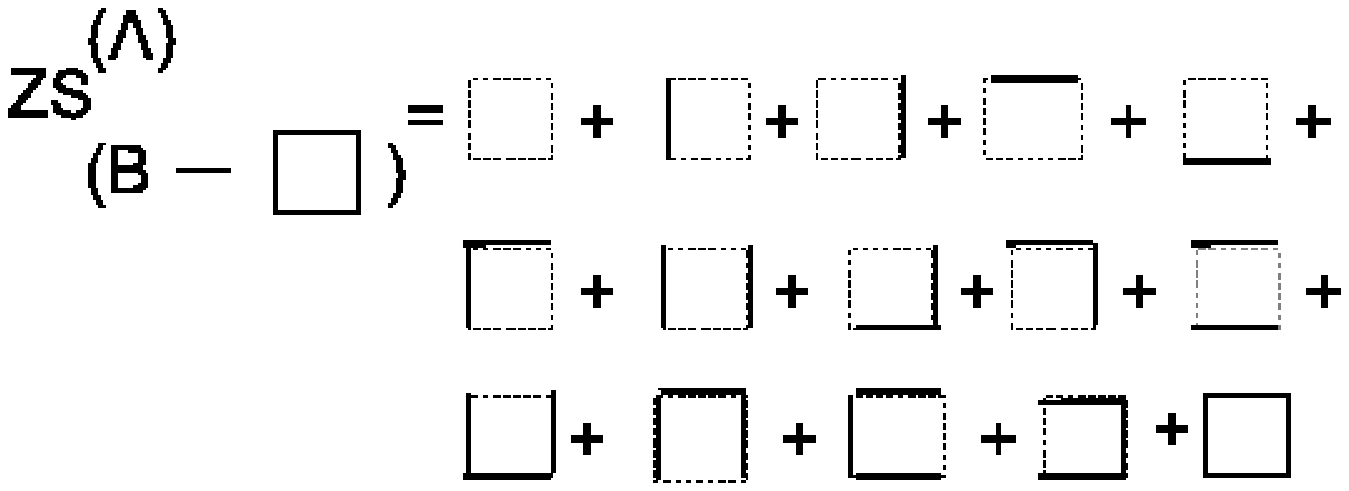}
\end{equation}

In the r.h.s we find terms labeled by paths arising after removing certain segments in the square. For instance, the label in the second square of the r.h.s indicates that in this term the Gaussian measure has Dirichlet boundary condition in all the lattice grid with the exception
of a single lattice segment $\mid$ (in bold), in which a value of $s$ between $0$ and $1$ is allowed.

As in the case of the function of two variables, each term is defined in terms of the Schwinger functions (depending on $s$) and integrals of certain partial derivatives with respect to  the $s_i$. Their precise definition is not relevant for our schematic presentation.

At least formally, we can extend  this expansion for the Schwinger function with free boundary condition. For such fully coupled function, the expansion contains an infinite number of terms, labeled by all the possible subset of paths $\Gamma$ (of finite length) in $B$:

\begin{equation}
S^{\Lambda}_{\text{Free BC}}=\sum_{\Gamma} \text{Terms with Dirichlet BC in $B-\Gamma$}\label{cluster}
\end{equation}

The issue of the convergence of this series will be considered later. The term labeled by ${B-\Gamma}$ corresponds to the case in which unit squares are decoupled from each other with the exception of those having contact with $\Gamma$. See Fig 3. As far as the size of $\Gamma$ increase, the Gaussian measure used in the terms labelled by $B-\Gamma$ approaches to the free measure.

\begin{figure}[ht]
 \includegraphics{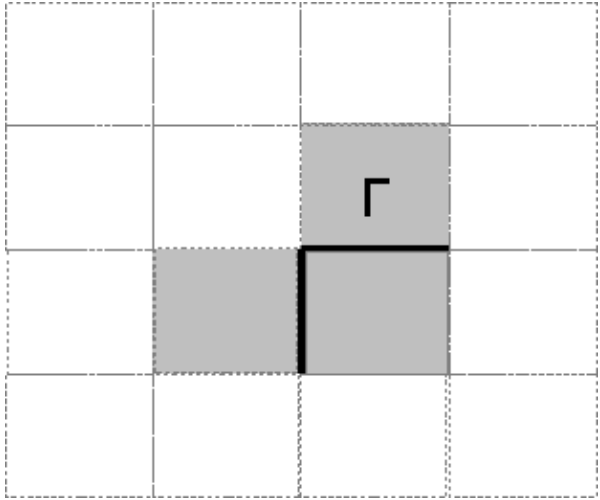}
\caption{The covariance has  Dirichlet BC in the set of all lattice lines (indicated by dotted lines) with the exception of $\Gamma$. Then, the only unit squares which are coupled are those which are connected through $\Gamma$. In the cluster expansion of the Schwinger functions, terms labeled by $\Gamma$ describe an almost decoupled quantity, in which the clusters of mutually decoupled regions are these shadow squares and the remaining unit lattice squares.}
\end{figure}

\item {\bf Factorization of each term of the series and resummation}

Each term in this expansion is almost decoupled, because there is only a finite number of segments in which Dirichlet conditions have not been imposed. For a given $\Gamma$, the term in the r.h.s of Eq. \ref{cluster} is constructed by the use of a Gaussian measure in which certain union of unit squares are coupled as it is showed in Fig.3. The coupled squares are those which share a unit segment belonging to the path $\Gamma$. Let us call them {\it clusters}.

So, each choice of $\Gamma$ determines a decomposition of $R^2$ into several clusters $X_i$, such that $C_{B-\Gamma}(x,y)=0$ for $x$ and $y$ belonging to different $X_i$'s. Therefore, each of these terms will inherit the factorization property we have mentioned.

After this observation, the next step consists in a convenient reorganization of the series by the use of the the factorization and a resummation. In order to explain that, let us consider for simplicity the particular case in which the arguments of the Schwinger function $x_1,x_2,...,x_n$ are contained in a single unit lattice square. The idea is to reorganize the sum in Eq. \ref{cluster} by considering a fixed cluster $X$ containing $x_1,x_2,...,x_n$ and collecting all the terms labelled by the collections of paths $\Gamma^{a}_{X}$  inducing a cluster decomposition containing the given $X$. There are several of such $\Gamma^{a}_X$. That is why we have introduced the label $a$.

In each term labeled by $X$, we can perform a factorization, being $T_X$ the factor associated to the cluster $X$. This factor will be the only one containing the information about the points $x_1,....x_n$. After doing that, we can make the sum of these terms (running over all the $\Gamma^{a}_X$'s) by using Eq. \ref{cluster} in the opposite sense. After doing that, we find the following expression:

\begin{equation}
S^{(\Lambda)}_{Free BC}=\sum_X  T_X \frac{\int e^{-\mathfrak{A}^{\Lambda-X}}d\mu^{B-X}}
{\int e^{-\mathfrak{A}^{\Lambda}}d\mu^{Free}}
\end{equation}

As we can see, the second factor (corresponding to the cluster $R^2-X$) is rather simple: it is $\frac{\int e^{-\mathfrak{A}^{\Lambda-X}}d\mu^{B-X}}
{\int e^{-\mathfrak{A}^{\Lambda}}d\mu^{Free}}$, the ratio of (cut-off) partition functions. One corresponds to the free BC and the other corresponds to the Dirichlet condition on ${B-X}$.

Despite the simplicity of the second factor, it seems that we have not gain so much, because there we have still the free boundary condition in the r.h.s in the denominator of the second factor. However, we should take into account that it appears in the ratio $\frac{\int e^{-\mathfrak{A}^{\Lambda-X}}d\mu^{B-X}}
{\int e^{-\mathfrak{A}^{\Lambda}}d\mu^{Free}}$. When $\Lambda\nearrow{R^2}$, the numerator approaches to the denominator, which left us with a limit of the type $\frac{\infty}{\infty}$.

After doing this factorization and resummation, we have the cluster expansion. It is a sum of $s$-derivatives of the Schwinger function with a Gaussian measure which establish a coupling among those unit squares contained in the cluster $X$ meeting the points $x_1,x_2,...x_n$. In each term, the size of $X$ is finite.

\item {\bf Convergence uniform with the size of $\Lambda$}

It can be shown that this cluster expansion converges for any $\Lambda$ of finite size. However, we are interested in the limit $\Lambda\nearrow{R^2}$. The terms of the series which are relevant for the infinite volume limit are the ones in which the size of the relevant clusters $X$ is large.

In this stage, it is more difficult to provide heuristic arguments. Such would require from us a deeper understanding of the nature of this proof. We will only say that after establishing several inequalities for both factors in the cluster expansion ($T_X$ and  $\frac{\int e^{-\mathfrak{A}^{\Lambda-X}}d\mu^{B-X}}
{\int e^{-\mathfrak{A}^{\Lambda}}d\mu^{Free}}$) it has been derived the following inequality for $\lambda/m^2$ small enough:

\begin{equation}
\sum_{\mid\text{lattice region X}\mid> D}\text{Terms with coupling in the lattice region $X$}<e^{-cD}
\end{equation}

being $c$ a constant which does not depend on $D$ and $\Lambda$.

That bound expresses that the previous series converge uniformly with the size of $\Lambda$. That implies that this series converge in the infinite volume limit.

Due to the omission of the technical details of the previous proof, it is difficult to explain here why it is required that $\lambda/m^2$ should be small enough. However, we want to point out that this is a different condition than the one appearing in the perturbative approach. We should take into account that the cluster expansion is not a sort of Taylor expansion in powers of the coupling constant. Roughly speaking, it is rather an expansion in the size $\mid{X}\mid$ of the  clusters containing the n-point of the Schwinger functions.

The convergence of the cluster expansion holds for any value of $\lambda/m^2<\epsilon$, being $\epsilon$ certain positive number. The convergence is not asymptotic in $\lambda/m^2$ as it occurs in the perturbatives series. It means that for each small enough value of $\lambda/m^2$ the cluster expansion defines the exact Schwinger functions.
\end{enumerate}

\section{Verifying that the resulting n-point functions come from correlation functions of a QFT}

It could seem that this procedure is never ending, because we still have to prove that the limiting Schwinger functions fulfill certain requirements: the OS axioms. However, what makes the Euclidean Gell-Mann-Low ansatz a convenient recipe is the fact that part of the OS axioms are fulfilled in the cut-off version (in a manifested way) and then these axioms are automatically verified in the infinite volume limit. The remaining one -cluster property- requires a more difficult proof.

\subsection*{Properties manifested in the Gell-Mann-Low ansatz}

Belonging to the first case, we have the following OS axioms:

\begin{itemize}
\item Symmetry
\item Euclidean invariance
\item Reflection positivity
\end{itemize}

\underline{Symmetry}: The more evident property is the symmetry of the Schwinger functions under permutation of the argument of the Gaussian fields appearing in the product. This property is naturally preserved under the infinite volume limit.

\underline{Euclidean invariance}: This property  is of a different nature, because it is not present in the cut-off version but only after the infinite volume limit is taken.
It is enough for our purpose to see the plausibility of such property: because the finite size of the region is the responsible of the breaking of the Euclidean invariance, it is natural to expect that this symmetry is restore once the cut-off region is extended to infinity.

\underline{RP}: that is trivially accomplished (in the cut-off theory) if the chosen cut-off region $\Lambda$ is invariant under time reflection. Decomposing $\Lambda$ as $\Lambda_+ + \Lambda_{-}$ (being $\Lambda_+$ and $\Lambda_{-}$ the positive and negative time hyperplane respectively), that means
 that $\Theta(\Lambda_{\pm})=(\Lambda_{\mp})$. RP can be seen if we rewrite the perturbed measure as:

 \begin{eqnarray}
 d\mu^{\text{non-Gaussian}}\equiv\frac{e^{-\frak{A}^{\Lambda}}d\mu}{Z^{\Lambda}}=\\\nonumber
 \frac{\theta(e^{-\frak{A}^{\Lambda_+}})e^{-\frak{A}^{\Lambda_+}}d\mu}{Z^{\Lambda}}
 \end{eqnarray}

Using this expression for the perturbed measure, RP in the form of Eq. \ref{RPutil} follows by writing:

\begin{equation}
\int\theta(A^{(n)})A^{(n)}d\mu^{\text{non-Gaussian}}=\int\theta(e^{-\frak{A}^{\Lambda_+}}A^{(n)})e^{-\frak{A}^{\Lambda_+}}A^{(n)}d\mu\geq0
\end{equation}

Because we are interested in the limit  $\Lambda\nearrow{R}^2$, we can take this limit by increasing the size of $\Lambda$, keeping its invariance under time reflection.

\subsection*{Clustering in the infinite volume limit}

Proving clustering property in the infinite volume limit presents a difficulty comparable to the case of cut-off version. That is because in the intermediate steps of the cluster expansion appear bounds for the Schwinger functios which are uniform with the cut-off volume. Such are used for the clustering property. We recommend the reading of \cite{Spencer2}, which contains an account of this step. That reference is more pedagogical than the original research paper \cite{Spencer}.

It is reasonable to expect a close relation between the clustering property and the finiteness of the Schwinger function. We have seen that if we use the completely decoupled measure, then it follows both the finiteness of the infinite volume and the clustering. We have already explain (intuitively) this link in section \hyperref[volumeninfinito]{\ref{volumeninfinito}}. That intuition is proved to be right in the case several case, including $\lambda\Phi^4$.

\subsection*{Remarks on the small size of $\lambda/m^2$}

We need to say something about the requirement on the constant $\frac{\lambda}{m^2}$: when it is said that it should be weak, that means that it belong to certain interval $[0,\epsilon]$. For each finite value in that interval, we have a non-perturbative description of the theory and not a mere asymptotic expansion in the coupling $\frac{\lambda}{m^2}$. That statement is different from the one of perturbative theory because the last one involves asymptotic series which does not converge for any small finite value of
the coupling constant.

Besides that, we want to point out that the smallness of the coupling constant is not a general condition used in the CQFT approach. As we have mentioned, there are other methods for controlling the infinite volume limit apart of the cluster expansion. One of these methods has been applied to polynomials of even degree plus a non-zero linear term (hence, excluding $\lambda\Phi^4$), showing that these models fulfill the OS axioms for any value of the coupling constant \cite{Simon}. When this method is applied to $\lambda\Phi^4$, it can be proven the existence of the infinite volume limit fulfilling all the OS with the exception of the cluster property. That means that this method was not successful in proving the uniqueness of the vacuum.

For $\frac{\lambda}{m^2}>>1$ (strong coupling) it can be shown by others means that there exist a decent quantum field theory. The description of this regime is beyond the scope of this note. We just mention this phenomena in order
to exemplify the existence of a difference between weak and strong regimes of a given theory. Here the expressions {\it weak} and {\it strong} have a literal meaning, being both applied to an existent QFT described in a non-perturbative way.

\section{Particle interpretation and new information beyond perturbative level}

The previous steps show that there exist a QFT fulfilling all the physical requirement encoded in the Wightman axioms. We called $\lambda\Phi^4$ to that theory, because the polynomial $F(x)=\lambda{x}^4$ was the term used for the perturbation of the Gaussian measure. It is natural to ask whether this theory will describe a quantum theory of interacting particles of spin zero with a $\lambda\Phi^4$ interacting term.

A particle interpretation is guarantied if the theory fulfills additional requirements, which were stated in an important theorem due to the successive work of Haag and Ruelle \cite{Ruelle}. Although the proof of that theorem is complicated (that is beyond the scope of thus note), the hypothesis in which the theorem is based on can be expressed in a very simple way: the mass operator $\hat{M}$ should have an
spectrum with a gap between $0$ and a positive value $M$.

That hypothesis is sufficient for the construction of certain states having the same behavior as the ones of the free theory. These states are constructed by the application of the field operator $\hat{\Phi}(h_M)$ to the vacuum, using an special set of test function $h_M$. Such function are chosen in such a way that the spectrum $\hat{M}$ on these states is the same that the one in a QFT of a free scalar field of mass $M$.

The proof of the existence of this gap in the mass spectrum can be obtained from an stronger version of clustering property than the one necessary for the existence of the QFT.

\subsection*{The mass gap}

If we read the complete proof of the cluster property in the case of the $\lambda\Phi^4$, we will see that it contains at the same time the proof that -for a weak value of the constant $\frac{\lambda}{m^2}$- the mass operator has no other eigenvalue in the interval $(0,M+\epsilon)$ than $0$ and $M$. So, according with the
Haag-Ruelle theory, we will have a particle interpretation for the asymptotic states, correspondent to scalar field of mass $M$.

In \cite{Epstein} it was also found a close expression for the physical mass $M$ of the asymptotic states, as a function of the parameters $\lambda$ and $m$ arising in the interacting term.

\subsection*{Bound states}

Having a non-perturbative definition of QFT with a particle interpretation, we are left with the difficult task of extracting practical physical information about the models. That is a better situation than the one of the perturbative description, because at the end is a computation issue.

One of the relevant issues is the existence of bound states in the model. That existence is also related to the properties of the spectrum of the mass operator. The
existence of a two particle bound state amounts to the existence of an eigenvalue in the interval $(M, 2M)$. That is the definition of what is a bound state, because the
eigenvalue $2M$ correspond to a two-particle asymptotic state. That definition captures the classical feature of a bound state: the lower energy of this state in comparison with the one of a the one composed by 2 free particles.

In the case of the theory of this note, it was proved that there are not 2-particle bound state. We have included this very incomplete description of this aspect of the model  in order to emphasize that the achievement of CQFT goes beyond the proof of the existence of models. Because the strategy were based on intuitive ideas, the CQFT approach is also able to extract physical relevant information.

\newpage

\part{Link with Hamiltonian approach and perturbation theory}
\newpage

\section{Hamiltonian point of view: Schwinger n-point functions as vacuum expectation value of interacting fields}

In the previous part of this note, we have followed the functional point of view, in which it was not required an explicit construction of the Hamiltonian operator and the interacting field. These objects are implicity defined by the reconstruction theorem we have mentioned at the
beginning. The link between the two descriptions can be formally written as:

\begin{equation}
\boxed{
(\Omega^{int}, \hat{\Phi}^{int}_{it_1}(x_1)..\hat{\Phi}^{int}_{it_n}(x_n)\Omega^{int})=
\lim_{T\rightarrow\infty}\frac{E(\Phi_{t_1}(x_1)...\Phi_{t_n}(x_n)e^{-\int_{-T}^{T}V(\Phi_t)dt})}{E(e^{-\int_{-T}^{T}V(\Phi_t)dt})}
}
 \end{equation}

where $V(\Phi_t)$ is the spatial integral of the term $:\Phi^4:(x,t)$. In this context we will make an explicit distinction between space and time; now, $x$ stands for the spatial coordinate. This is the $D=1+1$ version of the Eq.\ref{EGL}. Until now, we have described the different steps towards the definition of the r.h.s. The reconstruction theorem guaranties that each ingredient in the l.h.s exists.

In the following sections, we want to say something about the definition of $\hat{\Phi}^{int}$ and $\Omega^{int}$.
As we will see, the difficulties for making sense out of the Euclidean path integral has a counterpart in the difficulties for defining the interacting field in a Poincare invariant way. The Haag theorem (see \cite{Fraser} for a nice account), formulated in the middle of '55, shows that this difficulty is something general and not tied with a particular interaction term.

\section{Dealing with the three divergences in the Hamiltonian approach}
\subsection{Dealing with the first divergence: the definition of the interacting term}
We will consider the first divergence from the Hamiltonian point of view. But first, we need some preliminary notions arising in the Hamiltonian formalism.

\subsubsection{Operator, bilinear forms and fixed time operators}
{\bf Creation `operators' as bilinear forms}
\\

It is important to recall an elementary fact about operators and bilinear forms in a Hilbert space: {\it an operator always defines a bilinear form but there are bilinear forms which do not come from an operator}. Let us consider this statement with more detail. If we have
an operator $\hat{A}$, we can define an associated bilinear form $A$ by:

\begin{equation}
A(v,w)\equiv{<}v,\hat{A}w>
\end{equation}

being $v,w$ any vector of the Hilbert space. However, if we have a bilinear form $B$, there is not guaranty that there exists an operator $\hat{B}$ such that: $B(v,w)=(v,\hat{B}w)$. A relevant example of the last case is the so-called {\it creation operator} $a^{\dag}_k$ associated to a defined spatial momentum $k$. See pages 218,219 of \cite{Simon2} for a more extended explanation of the following.

We have already mentioned that the expression $\Phi(x)$ should be considered
as a formal expression and not as an operator. Then, it seems that the same applies to its decompositions of its formal Fourier transforms: the $a^{\dag}_k$ and $a_k$. However, it turns out that $a_k$ has a better
behavior than  its partner $a^{\dag}_k$: it can be considered as an operator, defined by the usual action on the Fock space. That follows by looking at the action of the
annihilation operator, which does not introduce singular expressions like $\delta(k)$ when is acting in the Fock space.

The status of $a_k$ as an operator makes possible a natural interpretation of {\it $a^{\dag}_k$ as a bilinear form} $A_k$ defined by:

\begin{equation}
A_k(v,w)\equiv<a_kv,w>
\end{equation}

This definition is motivated by the formal manipulation of $a^{\dag}_k$ as it were an adjoint operator of $a_k$:
$<a_k^{\dag}v,w>=<v,a_kw>$.

The same definition can be applied to expressions like ${({a^{\dag}}_{k'})}^n{(a_k)}^m$. It can be interpreted as a bilinear form $A_{k',k}$ defined by:

\begin{equation}
A_{k,k'}(v,w)\equiv{<}{a_{k'}}^nv,{a_k}^m w>
\end{equation}

{\bf $:{\hat{\Phi}(x,t)}^n:$ as a bilinear forms}
\\

We have said that $\hat{\Phi}(x,t)$ is just a formal expression, which is motivated by the consideration of $\hat\Phi(.)$ as regular operator valued distribution admitting a kernel: $\hat{\Phi}(f)=\int \hat{\Phi}(x,t)f(x,t)dxdt$. However, the previous observation concerning the status of $a^{\dag}_k$ as bilinear form shows that $\hat{\Phi}(x,t)$ can be also interpreted as a bilinear form.

If we still denote by $\hat\Phi(f)$ the bilinear form associated to the operator $\hat\Phi(f)$, we can get that
the equation $\hat{\Phi}(f)=\int \hat{\Phi}(x,t)f(x,t)dxdt$ is not merely a formal relation between operator but a meaningful equality between bilinear forms.

  Moreover, $:(\hat{\Phi}(x,t))^n:$ can also be interpreted as bilinear forms. By expressing $\hat{\Phi}(x,t)$ in terms of the $a^{\dag}_k$ and $a_k$, and using the definition of the normal order $::$ , we can see that $:(\hat{\Phi}(x,t))^n:$ is a sum of terms of the form:

\begin{equation}
:(\hat{\Phi}(x,t))^n:=\sum_{a=0}^{n}\int F_a(k_1,...k_n) a^{\dag}_{k_1}...a^{\dag}_{k_{a}} a_{k_{a+1}}...a_{k_n}dk_1dk_2..dk_n
\end{equation}

It can be checked the functions $F_a$ depend on the $k$'s in such a way that $:(\hat{\Phi}(x,t))^n:$ is a well defined bilinear form in the Hilbert space.

We want to emphasize that the status  $:(\hat{\Phi}(x,t))^n:$ as bilinear form (which holds for any spacetime dimension) does not imply that $:(\hat{\Phi}(x,t))^n:$ comes from an operator. However, in the particular case of $D=1+1$ the previous bilinear comes from an operator, which was called in this note $:{\hat\Phi}^n:$. This fact can be expressed in the following relation:

\begin{equation}
:{\hat\Phi}^n:(x,t)=\;:(\hat{\Phi}(x,t))^n:
\end{equation}

Again, the statement should be read as an equality between bilinear forms. As we have said, there is a minor difference in the notation: in the l.h.s, we want to stress that there exists an operator valued distribution $:{\hat\Phi}^n:$, whose associated bilinear form is the one given by the r.h.s.
\\

{\bf Fixed time Wick powers}
\\

Let us recall that one of the properties of the free field is the existence of fixed time $\hat{\Phi}_t$. That means that in $D=1+1$, $\hat{\Phi}_t$  is a distribution on the space of functions of a single variable. So, it makes sense expressions like $\int \Phi_t(x)g(x)$, being $g$ a function in $S(R)$. The relation between the operator value distribution in $S(R^2)$ and the fixed time version is the following:

\begin{equation}
{\Phi}(f)=\int \Phi_t(f_t) dt
\end{equation}

being $f$ a function of the space time and $f_t$ the function of a single variable given by: $f_t(x)=f(x,t)$.

We can ask if there exist a fixed time version of a Wick power. Such will be the first step for the definition of an interacting Hamiltonian density. A consideration made in \cite{Jaffe3} shows that this is the case for the operator valued distribution $\hat{\Phi}^n$. So, it makes sense the expression $:\hat{\Phi}^n_t:$ as a distribution acting on functions of a single variable.

The existence of the fixed time operator valued distribution $:\hat{\Phi}^4:$ is necessary for the definition of the interacting term.
\\

{\bf The particular case of the free Hamiltonian}
\\

The Hamiltonian $H_0$ of the free field is defined as the infinitesimal generator of the translation. Its action on the Fock space can be
easily written. For the the case of a 1-particle state, given by the function $\Psi$ of the spatial momentum, the action of the $H_0$ is:

\begin{equation}
(H_0\Psi)(k)=i\omega(k)\Psi(k)
\end{equation}

being $\omega(k)=\sqrt{k^2+m^2}$. There is an analogous expression for the action of $H_0$ in a general $n$-particle state.

As we see, the definition of the Hamiltonian does not require to write any expression involving the free field. However, we are familiar with the following expression for the Hamiltonian:

\begin{equation}
H_{free}=\frac{1}{2}\int  :{(\hat{\Pi}_0(x))}^2 + {(\partial_x\hat{\Phi}_0(x))}^2 + m^2{(\hat{\Phi}_0(x))}^2:dx\label{freeH}
\end{equation}

being $\hat{\Pi}_0(x)$ the temporal derivative of $\hat{\Phi}_0(x)$.

This equality makes sense as a statement about bilinear forms. The l.h.s should be understood as the {\it bilinear form associated to the free Hamiltonian} . The r.h.s is already a bilinear form.

Let us remark that in the r.h.s can be considering as the result of smearing the expression $:{(\hat{\Pi}_0(x))}^2 + {(\partial_x\hat{\Phi}_0(x))}^2 + m^2{(\hat{\Phi}_0(x))}^2:$ with the the constant function equal to 1. The fact that the final result comes
from an operator is an special case. In other cases, like $:\hat{\Phi}_0^4:$, we can not expect that  $\int :\hat{\Phi}^4_0(x):dx$ makes sense.

\subsubsection{The canonical quantization with the cut-off interacting term}

According with the last remark, the expression ${:}\Phi_0^4:(g)$ makes sense if we choose $g$ as a function in the space $S(R)$, which can not be a constant. We can take $g$ as the characteristic function with value
$\lambda$ in the interval of size $L$\footnote{The characteristic function is not differentiable every where. So, it is not a allowed as test function in $S(R)$. More properly, we should consider an smooth function in $S(R)$ with support on the region of size $L$. That remark is not relevant for the following discussion.}. This choice of $g$ leads to {\it spatial cut-off Hamiltonian interacting term}, which formally could be expressed as $\int_0^L  :\hat\Phi^4_0:(x) dx$. We will denote it as $V_L$, in order to stress that the interaction is turned on only on the region of size $L$ for any time.

The interacting cut-off field, in the Heisenberg representation, is defined by:

\begin{equation}
\hat{\Phi}^{L}(t,x)=e^{i(H_0+V_L)t}\hat{\Phi_0}(x)e^{-i(H_0+V_L)t}\label{cutoffevolution}
\end{equation}

Motivated by the particular case of the free Hamiltonian (which can be expressed as an integral without cut-off in Eq. \ref{freeH}) we can take the limit in which the characteristic function approaches the constant function with value $\lambda$. The existence of that limit would correspond to the definition of a Hamiltonian without cut-off acting in the Fock space. As we will see, that is not a trivial issue.

\subsection{Dealing with the second divergence: the stability proof}

Now, we want to see at a heuristic level the relation between the integrability condition $E(e^{\mathfrak{A}^{\Lambda}})$ and the boundedness from below of the cut-off Hamiltonian $H_L\equiv{H}_0+V_L$. Let us start considering the Hamiltonian $H_L^{\kappa}=H_0 + V^{\kappa}_L$, being the interacting term defined by: $V_L^{\kappa}=\int_L :{\hat{\Phi}_{0,\kappa}}^4:(x)$ and let us assume that the interacting term fulfills the suitable conditions in order that the following Feynman-Kac formula hold:

\begin{equation}
(\Omega,e^{-TV_L^{(\kappa)}}\Omega)=E(e^{-\int_0^T V_L^{(\kappa)}(\Phi_t)dt})
\end{equation}

This relation shows that {\it free vacuum expectation} of $e^{-TV_L^{(\kappa)}}$ is related to the Gaussian expectation  $E(e^{-\int_0^T V_L^{(\kappa)}(\Phi_t)dt})$. The last quantity is in fact the integral $\int e^{-\frak{A}_{\kappa}^{\Lambda}} d\mu$ if we choose the spacetime region $\Lambda$ as a rectangle of sizes $T$ and ${L}$.

Because $\Omega$ is the vacuum of the free theory, this formula is not useful for our purpose. What we need is an equality or inequality having in the l.h.s the interacting vacuum. There exists such inequality, whose derivation goes beyond the scope of this note. By taking the limit $\kappa\rightarrow\infty$ in such inequality, it can be derived the bound:

\begin{equation}
-E^L \leq\frac{1}{\alpha(T)}log E(e^{-\int_0^T V_L(\Phi_t)dt})\label{cutoffenergia}
\end{equation}

where $E_L$ is the negative lowest eigenvalue of the Hamiltonian $H_L$ and $\alpha(T)$ is a function of $T$ whose detailed expression is not relevant.

The inequality of Eq. \ref{cutoffenergia} shows that the finite integral $-log E(e^{-\int_0^T V_L(\Phi_t)dt})$ is a lower bound of the interacting Hamiltonian. This does not pretend to be a proof that the integrability condition of the exponential implies that the interacting Hamiltonian is bounded from below. See pages 158-161 of \cite{Simon} for more details.

\subsection{Dealing with the third divergence: the removal of the spatial cut-off}

 Let us consider again the case in which the region $\Lambda$ used in the Euclidean approach is is a rectangle of size $T$ and $L$. We want to see from the Hamiltonian point of view the different meaning of the limits $L\rightarrow\infty$ and $T\rightarrow\infty$, which were treated in an equal foot in the Euclidean approach.

As we have mentioned, the existence of the limit $L\rightarrow\infty$ for the exponential term amounts to the possibility of extending the scope of the test function to an infinite large spatial interval for $:\hat\Phi_0^4:$. That will be considered soon.

The limit $T\rightarrow\infty$, instead, is of a different nature. From the Hamiltonian point of view, we are not doing anything!. Let us recall the Gell-Mann-Low formula for quantum mechanics case:

\begin{equation}
(\Omega^{int}, \hat{\Phi}^{int}_{it_1}..\hat{\Phi}^{int}_{it_n}\Omega^{int})=
\lim_{T\rightarrow\infty}\frac{E(\Phi_{t_1}...\Phi_{t_n}e^{-\int_{-T}^{T}V(\Phi_t)})}{E(e^{-\int_{-T}^{T}V(\Phi_t)})}
\end{equation}

The limit $T\rightarrow\infty$ does not tell anything about the quantum mechanics system of the l.h.s. $T$ is just a parameter which is necessary in order to relate the interacting vacuum with the free vacuum. This relation in fact leads to the Gell-Mann-Low formula. In $D=1+1$, the Gell-Mann-Low formula take a similar form::

\begin{equation}
(\Omega^{int}, \hat{\Phi}^{int}_{it_1}(h_1)..\hat{\Phi}^{int}_{it_n}(h_n)\Omega^{int})=
\lim_{T\rightarrow\infty}\frac{E(\Phi_{t_1}(h_1)...\Phi_{t_n}(h_n)e^{-\int_{-T}^{T}V_L(\Phi_t)dt})}{E(e^{-\int_{-T}^{T}V_L(\Phi_t)dt})}
\end{equation}

So, if we look the r.h.s, we see that the contact between the Schwinger function with the spatial cut-off QFT vacuum correlation functions is established once `half' of the infinite volume limit is taken. Let us recall that what we have called infinite volume limit in the Euclidean approach was $\Lambda\nearrow{R}^2$ and not merely $\mid\Lambda\mid\rightarrow\infty$.

\subsubsection{The van-Hove phenomena and the Haag theorem}

The existence of the limit $L\rightarrow\infty$ has another meaning. A non trivial part of the construction of the model in the Hamiltonian approach is that of showing that the unitary evolution given by $H_L$ makes sense in the limit $L\rightarrow\infty$. That was shown in \cite{Jaffe2}. But there is an important remark concerning the meaning of the existence of the limit. Let see the behaviour of the vacuum state $\Omega^L$ and its eigenvalue $E_L$ of the cut-off Hamiltomnian in the limit $\rightarrow\infty$.

It can be shown that $\lim_{L\rightarrow\infty}E_L=-\infty$. Apart of this divergence, something strange happens with the overlap between the interacting vacuum $\Omega^L$ and the free vacuum $\Omega$. It can be proved that $(\Omega^L,\Omega)\leq{e^{-cL}}$, being $c$ a positive constant. From this, it follows that:

\begin{equation}
\lim_{L\rightarrow\infty}(\Omega^L,\Omega)=0
\end{equation}

Such is an example of the so-called {\it van Hove phenomena} (see \cite{Simon}, pag. 185). The name comes from an early observation by van Hove about this phenomena in certain QFT \cite{vanHove}.

That means that in the infinite volume limit, the limiting interacting vacuum state can not belong to the Hilbert space of the cut-off theory. That is: $\Omega^{int}\neq\lim_{L\rightarrow\infty}\Omega_L$. What can be proven is that the true vacuum lives in a non unitarily representation of the free theory. That makes the difference with the cut-off theory, in which it the interacting picture was used according with Eq. \ref{cutoffevolution}.

 The previous phenomena is a very pedagogical illustration of the {\it Haag theorem}, formulated in the middle of '50, which states that it is not possible to have a representation of canonical commutation relation (CCR) of an interacting theory which results to be unitarily equivalent to the CCR representation of the free field theory \cite{Fraser}. Such was possible for the cut-off version, because one key assumption of the Haag theorem was avoided: the translation invariance.

Haag theorem has a conceptual value for the understanding of the obstacles for the definition of a interacting QFT, clarifying the role of the cut-off. Most of the standard textbook does not take into account this obstruction when the Dyson operator is written,
assuming the existence of the interacting picture. Of course, this does not conduct to any wrong statement because this step pretend to be only an heuristic guide for the derivation of the perturbative series. The Haag theorem recall us that the perturbative series has not been deduced from the meaningless non-perturbative expression but that the series constitutes the (perturbative) definition of the QFT.

This subtle is not manifest in the Euclidean approach as far as we focus in the correlation functions themselves and not in the reconstruction of the QFT from which these come from.

\section{Perturbative series and the exact $n$-point functions}

The natural question after this long construction is: how is the model of this note related to the  perturbative $\lambda\Phi^4_2$?.

Before answering such question, we want to remark that the absence of a relation between both would not invalidate the previous construction. The model we have considered fulfills all the physical requirements of a relativistic quantum field theory. The agreement with the perturbative treatment is not required by the GW axioms. However, such a link would be desirable, because at the end we want to find a non-perturbative version of realistic QFT -checked in the laboratory- which are formulated perturbatively.

\subsection{From the exact n-point functions to its Taylor series}

It is natural to expect that the series arising by making a Taylor expansion of the n-point functions agree with the usual one. That is because the standard derivation of the perturbative series starts from the formal non-perturbative expression which was shown in the CQFT to be well defined.

Let us go to the issue of the the convergence of this series. We know from \cite{Jaffediv} that the series of $\lambda\Phi^4$ are not convergent in the standard sense. However, in \cite{Dimock1} was proven that for the general case of  all polynomial bounded from below $P(\Phi)_2$ the perturbative series of the Schwinger functions are {\it asymptotic} to the non-perturbative expression. Moreover, this result was extended in \cite{Frolich} to the case of the perturbative series of the $S$ matrix \footnote{Let us recall that the $S$-matrix is defined in terms of {\it time ordered} n-point functions. This difference introduce further complications}.

In addition to the asymptotic convergence, it was proven in \cite{Eckmann} that the asymptotic series converge in a {\it Borel sense} to the exact Schwinger functions corresponding to a polynomial of order 4 interactions.

\subsection{The proper use of the asymptotic convergence of the perturbative series}

Now, we have the full n-point functions (fulfilling the general requirement of a relativistic quantum field theory) having the perturbatives series as their asymptotic expansion. The asymptotic convergence has a practical value: this ensures
that the difference between the truncated series-at order $N$-and the value of the n-point functions will be of order $\lambda^{N+1}$. The agreement with the full n-point functions will be better as far as $\lambda$ goes to zero. That is why these series are
reliable at weak coupling.

We want to stress that the practical value of the series is based on {\it the existence of the full n-point functions to which these approach}. Of course, that existence is the implicit assumption which justify that physicists confront the truncated series with the experience.

However, the asymptotic convergence can not help to {\it define} the n-point function. That is because there is not a unique function having a given series as its asymptotic expansion. We want to emphasized that this is true for any finite range of the coupling constant, no matter how small is. That is because the radius of convergence of the series is not small but zero.

We have followed the common distinction between perturbative and non-perturbative approach although we do not consider that very appropriated: that terminology has an attenuating effect, suggesting that the difference merely regards the regime in which the theory is described. However, when it is said that the perturbative approach describe the theory at weak coupling, we should have the previous observation in mind. That lead us to appreciate that the role of the construction of the non-perturbative n-point function of $\lambda\Phi^4$ is not that of extending the regime of the perturbative theory.

\section{The three main risks of divergences in the perturbative approach: where are they?}


If we compare the CQFT and the perturbative approach (both the Hamiltonian and the functional approaches) to $\lambda\Phi^4$ we will find important differences. The main one is the absence of the risk of divergence in step II and III. The reason for that difference is trivial: the difficulties in
the step II and III are associated with the introduction of an exponential of the interaction. So, these difficulties are reduced (some of them are eliminated) when the exponential is expanded as a formal Taylor series.

\subsection{Dealing with the first divergence: the trivial renormalization and the ultraviolet divergences}

Most of the expositions of the perturbative approach start with the infinite volume interacting term. However, in order to see the analogous of the first risk of divergence, we should consider a {\it cut-off perturbation theory}.

This first step is the only one appearing explicitly in the perturbative approach. It appears in the regularization procedure. In the particular case of $\lambda\Phi^4$ we have not ultraviolet divergence. However, there are diverges that are eliminated by the introduction of the Wick order of the fourth power of the field. Usually, this trivial step is not considered as part of the regularization. The Wick order in fact eliminates divergences contained in the so-called {\it tadpole} diagrams.

Let us point out a minor difference between the way in which this step is presented in each approach. For simplicity, let us consider both the non-perturbative and perturbative Hamiltonian approaches. In the first case, this step is done in order to make the interacting term a {\it well defined operator}. In the perturbative approach, the regularization consists in making {\it well defined the expectation values} involving the interacting term. In other words, in the CQF approach the regularization is done in the beginning, guarantying that any expectation value involving the interacting term makes sense.

In the particular case of $\lambda\Phi^4$ in $D=1+1$ in turns out that the normal order is enough for making the interacting term a well defined operator. However, this is not mentioned explicitly in the perturbative approach.

\subsection{The boundedness from below of the Hamiltonian: why we do not see this issue in the perturbative approach?}

The series in the functional perturbative approach are defined by a Taylor expansion of a formal exponential expression. Without exponential there is not any risk of this type of divergence. The task of regularization consists merely in making finite each term of the expansion. Of course, from the finiteness of the each term of the Taylor expansion we could not conclude the finiteness of the integral of the exponential (at the end, this is related to the divergent character of the series).

We have mentioned that the integrability of the exponential in the Euclidean amounts to the boundedness from below of the Hamiltonian. In the Hamiltonian perturbative approach we do not see such a problem because {\it the series are not used for the computation of the full interacting Hamiltonian}. Such is only a formal expression which is written in the Dyson operator at the beginning of the procedure.

\subsection{The infinite volume and the cancellation of infinities in the perturbative approach}

Again, because the perturbative approach is not worried about the exact n-point functions, it is never considered the most complicated part of the non-perturbative approach: the proof of the convergence of the Schwinger functions in the infinite volume limit.

However, we can see a signal of this divergence in a detail of the procedure used for the definition of perturbative series. Let consider the Taylor expansion of the Schwinger functions with a cut-off in a region $\Lambda$. We will find different Wick contractions in each term. Among them, we will find:

\begin{enumerate}
\item $E(\Phi(x_1)\Phi(x_2)\mathfrak{A}^{\Lambda}\mathfrak{A}^{\Lambda})$
\item $E(\mathfrak{A}^{\Lambda}\mathfrak{A}^{\Lambda})$
\end{enumerate}

The first terms are finite even in the infinite volume limit. These terms correspond to integrals which have not ultraviolet divergence in $D=1+1$.

The second terms correspond to the so-called `bubble diagrams´. These are finite when there is a cut-off interacting term. We have mentioned that expectation of Wick power of the field are under control. In the infinite volume limit, we can see that bubble diagrams like this $E(\mathfrak{A}^{\Lambda}\mathfrak{A}^{\Lambda})$ diverge.

In the standard perturbative approach, the starting point is the infinite volume limit of such expression. The reason why this divergence is declared harmless is because that arise both in the numerator and the denominator of the series for the n-point functions. This claim is improved when it is said that the n-point functions are defined by the truncated expansion, in which this diagrams are omitted.

Again, here enters the issue of the non-convergence of the series: if the perturbative series were convergent, the proof of the finiteness of the n-point functions, for $\Lambda\rightarrow\infty$, would be more easy, being reduced to checking the cancellation of bubble diagrams in the truncated series.

This is one example of what we have addressed since the beginning: some of the non-trivial obstacles toward a non-perturbative definition of a relativistic invariant QFT are not manifested in the perturbative approach. Here, we see how the most hard obstacle (the existence of the infinite volume limit) is reduced to a mere cancelations of divergent factors.

\subsubsection*{A minor comment on the abuse of language used within some expositions of the perturbative approach}

We want to make a minor comment on the way in which the cancellation of the infinite volume divergences is expressed in some expositions of the perturbative approach.

Let consider $A_{\kappa},B_{\kappa}$ functions of a variable $\kappa$  diverging for $\kappa\rightarrow\infty$ in such a way that the limit $\lim_{\kappa\rightarrow\infty}\frac{A_k}{B_k}$ exists. We are aware that it does not make sense the limit: $\frac{\lim_{\kappa\rightarrow\infty}{A_k}}{\lim_{\kappa\rightarrow\infty}{A_k}}$. That lead to a meaningless expression of the type $\frac{\infty}{\infty}$.

In the non-perturbative approach, we find such limit in the statement:
\begin{equation}
\lim_{\Lambda\rightarrow\infty}\frac{\int \Phi(x_1)...\Phi(x_n)e^{-\frak{A}^{\Lambda}}  d\mu}{\int e^{-\frak{A}^{\Lambda}}d\mu }<\infty
\end{equation}

We do not need to talk about `cancelation of infinities´ but the limit of a quotient whose numerator and denominator diverge when $\Lambda\rightarrow\infty$.

Because in most of the standard exposition of QFT the n-point functions are not being considered as a result of a limit of a spacial cut-off version, a similar well defined limit can not be written. Instead, it is simply declared that the series for the the n-point functions are defined by the truncated expansion, in which the ill defined bubble diagrams are omitted. This ad-hoc definition could be avoided by simply defining the infinite volume formal series as a limit of cut-off perturbation series.

\section{Concluding remarks}

\subsubsection*{The role of the rigour in the CQFT approach}
A frequent prejudice is that the merit of a rigourous mathematical approach to a physical theory is merely the justification of statements which were derived by heuristic arguments. Such a prejudice does not apply in the case of the CQFT approach.

We have seen that due the rigour of the approach meaningless expressions have acquired a precise meaning. Such is the case of Euclidean path integral in the infinite volume limit. These expressions
appear in the standard approach and are usually manipulated in a formal way. In those cases, the role of CQFT is not merely to justify statements involving this quantities; before having a meaning, these are not statements at all, but sequence of symbols waiting for a semantic meaning.

A similar idea was expressed by the mathematicians Kurt Friedrichs in his book {\it Mathematical Aspects of the Quantum Theory of Field}, in the years after the development of the perturbative approach:

\begin{quote}
    {\it ``It is difficult for a mathematician to gather such information by reading papers and books addressed to physicists. It is not at all lack of rigor in the mathematical deductions which creates the difficulty; it is rather that the mathematical terms employed are not always defined precisely and that often their physical significance is not explicitly explained"}
\end{quote}

\subsubsection*{Important issues missing in the perturbative approach}

After looking at this simple model, we understand why it is more easy to define a theory by its perturbative series. If we do not make the Taylor expansion, we are forced to make a
more careful study of the  interacting term. In fact, the main difficulties of the model we have considered were those related with the divergences involved in the step II and III. Such are related to the fact that we have an
exponential and not a power of the interacting term. The perturbative approach avoids the confrontation with this problem paying a high price: {\it the well defined series result to be divergent and as a consequence these can not define the wished n-point functions}.

\subsubsection*{The increasing difficulties in higher dimension}
Because this is not a review on the status of CQFT, we have not considered more complicated models like $\lambda\Phi^4$ in $D=2+1$. However, we want to make a brief mention of new aspects arising in $D>2$. In $D=2+1$ the ultraviolet divergences in the
perturbative approach have also a counterpart in the step I towards the definition of the Schwinger function: the normal order is not enough to define the interacting term. We should write a modified polynomial, involving no-linear terms in the
coupling constant. Due to this apparent minor change, all the subsequent steps become more complicated. The correspondent increment in the difficulties is manifested in the reduced amount of successful models in comparison with $D=2$.

$\lambda\Phi^4$ in $D=3+1$ is a more complicated case, which has not been yet quantized along this lines. Moreover, there is evidence that it is not possible to have an interacting theory correspondent to $\lambda\Phi^4$ along the lines we have mentioned (see \cite{Sokal}). At
the present we have not a single example of a non-perturbative interacting relativistic QFT in $D=3+1$. Such is an open problem, deserved to be solved.\\

\subsubsection*{The need of a simpler procedure}

Although we have emphasized that the ideas behind each step have a clear interpretation, it is also true that some steps (like the proof of existence of the infinite volume limit) become very complicated from the technical point of view. Such difficulties are peculiar to each model and there is not a general way to deal with these. In the opinion of the authors, due to the small size of the CQFT community little progress on the simplification of some of the steps has been done. Concerning the paper \cite{Jaffe4} in which was developed  $\lambda\Phi^4$ model in $D=2+1$, it was said: "Written almost thirty five years ago, that paper has not yet fully digested and should be investigated from a more modern perspective " \cite{Jaffe2}. That statement can be applied to any other model. In this direction there are some progress in \cite{Rivasseau}, in which is considered how to avoid ``painful" steps of the CQFT approach.\\

\subsection*{A guide for further readings}

As we have emphasized along the note, this is a oversimplified exposition of $\lambda\Phi^4$ in $D=1+1$. This simplification could give rise to misunderstandings. We are not worry about the {\it omission of several proofs} but  the lack of precise definitions for most of the objects and limits appearing in this note. For instance, when we have introduced the Schwinger functions, we have omitted a list of requirements concerning the decay and the singularities of these objects. Concerning the existence of the interacting term as a limit, we have not provided a precise indication of the way in which the limit should be taken. The amount of omissions of this type is more important in the issue of the infinite volume limit.

However, we expect that this note provides the skeleton of the construction of more complicated models and act as a guide for the reading of a rigorous exposition. In the opinion of the authors, a nice way to get deeper into the subject could start by reading the modern textbook \cite{Dimock}, in particular chapters 11-13. In this reference it is introduced the Gaussian processes description of QFT and the steps I and II towards the definition of the cut-off Schwinger functions. That is done in a more detailed way. A complementary reading could be the chapter I-V of the book \cite{Simon}. Each of these references would facilitate the reading of the book \cite{Jaffe1}, where it is also considered in more detail the infinite volume limit of a family of scalar field polynomial interaction, and there are also addressed advanced topics including the treatment of gauge theories.

{\bf Acknowledgements}

This work was supported by CONICET,ANPCyT and UBA. ML thanks for the hospitality of Institute of Theoretical Physics, Vienna University of Technology. We thank Sergio Yuhjtman for discussions and also Juliana Osorio and Gaston Giribet for his carefully reading of this note.
ML gives special thanks Arthur Jaffe, by being open to answer questions on the issue. That helped us to correct misunderstandings and identify the main ideas of the subject.

\end{document}